\documentclass[12pt]{article}

\usepackage{cite}
\usepackage{axodraw}
\usepackage{epsfig}

\makeatletter

\def\paragraph{\@startsection{paragraph}{4}{\z@}{+2.00ex plus
 +1ex minus +.2ex}{1.5ex plus .2ex}{\it\normalsize}}

\def\section{\@startsection {section}{1}{\z@}{+3.0ex plus +1ex minus
  +.2ex}{2.3ex plus .2ex}{\normalsize\bf\boldmath}}
\def\subsection{\@startsection{subsection}{2}{\z@}{+2.5ex plus +1ex
minus +.2ex}{1.5ex plus .2ex}{\normalsize\bf\boldmath}}
\def\subsubsection{\@startsection{subsubsection}{3}{\z@}{+3.25ex plus
 +1ex minus +.2ex}{1.5ex plus .2ex}{\normalsize\it}}

 \expandafter\ifx\csname mathrm\endcsname\relax\def\mathrm#1{{\rm #1}}\fi

\newcounter{saveeqn}

\@addtoreset{equation}{section}

\def\nl{\nonumber\\}

\newcommand{\lsim}
{\mathrel{\raisebox{-.3em}{$\stackrel{\displaystyle <}{\sim}$}}}

\def\asymp#1%
{\mathrel{\raisebox{-.4em}{$\widetilde{\scriptstyle #1}$}}}

\def\Nequal#1%
{\mathrel{\raisebox{-.5em}{$\stackrel{=}{\scriptstyle\rm#1}$}}}
\newcommand{\dsl}[1]{\not \hspace{-0.7mm}#1}
\def\dsl{\mathpalette\make@slash}
\def\make@slash#1#2{\setbox\z@\hbox{$#1#2$}%
  \hbox to 0pt{\hss$#1/$\hss\kern-\wd0}\box0}

\def\beq{\begin{equation}}
\def\eeq{\end{equation}}
\def\beqar{\begin{eqnarray}}
\def\eeqar{\end{eqnarray}}
\def\barr#1{\begin{array}{#1}}
\def\earr{\end{array}}
\def\bfi{\begin{figure}}
\def\efi{\end{figure}}
\def\btab{\begin{table}}
\def\etab{\end{table}}
\def\bce{\begin{center}}
\def\ece{\end{center}}
\def\nn{\nonumber}

\def\text{\textstyle}

\def\Ga{\Gamma}
\def\ga{\gamma}
\def\de{\delta}
\def\De{\Delta}
\def\eps{\epsilon}
\def\veps{\varepsilon}

\def\la{\lambda}

\def\si{\sigma}

\def\refeq#1{\mbox{(\ref{#1})}}

\def\reffi#1{\mbox{Figure~\ref{#1}}}

\def\refse#1{\mbox{Section~\ref{#1}}}

\def\refapp#1{\mbox{App.~\ref{#1}}}
\def\citere#1{\mbox{Ref.~\cite{#1}}}
\def\citeres#1{\mbox{Refs.~\cite{#1}}}

\newcommand{\TeV}{\unskip\,\mathrm{TeV}}
\newcommand{\GeV}{\unskip\,\mathrm{GeV}}
\newcommand{\MeV}{\unskip\,\mathrm{MeV}}

\newcommand{\fb}{\unskip\,\mathrm{fb}}

\newcommand{\rI}{{\mathrm{I}}}
\newcommand{\rd}{{\mathrm{d}}}

\newcommand{\rT}{{\mathrm{T}}}

\newcommand{\M}{{\cal{M}}}

\def\mathswitchr#1{\relax\ifmmode{\mathrm{#1}}\else$\mathrm{#1}$\fi}

\newcommand{\PW}{\mathswitchr W}

\newcommand{\Pg}{\mathswitchr g}

\newcommand{\PH}{\mathswitchr H}

\newcommand{\Pu}{\mathswitchr u}

\newcommand{\Ps}{\mathswitchr s}

\newcommand{\Pb}{\mathswitchr b}
\newcommand{\Pbbar}{\mathswitchr{\bar b}}
\newcommand{\Pp}{\mathswitchr p}
\newcommand{\Pt}{\mathswitchr t}
\newcommand{\Ptbar}{\mathswitchr{\bar t}}
\newcommand{\Pep}{\mathswitchr {e^+}}
\newcommand{\Pem}{\mathswitchr {e^-}}

\def\mathswitch#1{\relax\ifmmode#1\else$#1$\fi}

\newcommand{\Mb}{\mathswitch {m_\Pb}}
\newcommand{\Mt}{\mathswitch {m_\Pt}}
\newcommand{\Mg}{\mathswitch {m_\Pg}}

\def\solid{\raise.9mm\hbox{\protect\rule{1.1cm}{.2mm}}}
\def\dash{\raise.9mm\hbox{\protect\rule{2mm}{.2mm}}\hspace*{1mm}}

\def\ie{i.e.\ }
\def\eg{e.g.\ }

\newcommand{\LO}{{\mathrm{LO}}}

\newcommand{\virt}{{\mathrm{virt}}}
\newcommand{\ferm}{{\mathrm{ferm}}}
\newcommand{\bos}{{\mathrm{bos}}}

\def\Re{\mathop{\mathrm{Re}}\nolimits}

\def\lra{\mathop{\mathrm{\leftrightarrow}}\nolimits}

\newcommand{\papb}{(p_ap_b)}

\def\draftdate{\relax}
\def\mda{\relax}
\def\mua{\relax}
\def\mla{\relax}
\def\Mda{\relax}
\def\Mua{\relax}
\def\Mla{\relax}
\def\draft{
\def\thtystars{******************************}
\def\sixtystars{\thtystars\thtystars}
\typeout{}
\typeout{\sixtystars**}
\typeout{* Draft mode!
         For final version remove \protect\draft\space in source file *}
\typeout{\sixtystars**}
\typeout{}
\def\draftdate{\today}
\def\mua{\marginpar[\boldmath\hfil$\uparrow$]%
                   {\boldmath$\uparrow$\hfil}%
                    \typeout{marginpar: $\uparrow$}\ignorespaces}
\def\mda{\marginpar[\boldmath\hfil$\downarrow$]%
                   {\boldmath$\downarrow$\hfil}%
                    \typeout{marginpar: $\downarrow$}\ignorespaces}
\def\mla{\marginpar[\boldmath\hfil$\rightarrow$]%
                   {\boldmath$\leftarrow $\hfil}%
                    \typeout{marginpar: $\lra$}\ignorespaces}
\def\Mua{\marginpar[\boldmath\hfil$\Uparrow$]%
                   {\boldmath$\Uparrow$\hfil}%
                    \typeout{marginpar: $\uparrow$}\ignorespaces}
\def\Mda{\marginpar[\boldmath\hfil$\Downarrow$]%
                   {\boldmath$\Downarrow$\hfil}%
                    \typeout{marginpar: $\downarrow$}\ignorespaces}
\def\Mla{\marginpar[\boldmath\hfil$\Rightarrow$]%
                   {\boldmath$\Leftarrow $\hfil}%
                    \typeout{marginpar: $\lra$}\ignorespaces}
\overfullrule 5pt
\oddsidemargin -15mm
\marginparwidth 29mm
}

\def\stars{\strut\leaders\hbox{*}\hfill\strut}
\def\starline{\hfil\strut\hfil\hbox to \textwidth {\stars}\hfil}

\begin{document}
\enlargethispage{2cm}
\thispagestyle{empty}
\def\thefootnote{\fnsymbol{footnote}}
\setcounter{footnote}{1}
\null
\draftdate
\hfill KEK-CP-213\\
\strut\hfill KEK Preprint 2008-13\\
\strut\hfill MPP-2008-41\\
\strut\hfill PSI-PR-08-07\\
\vspace{1.5cm}
\begin{center}
{\Large \bf\boldmath
NLO QCD corrections to
$\Pt\bar\Pt\Pb\bar\Pb$ production
\\[.5cm]
at the LHC: 
1.~quark--antiquark annihilation
\par} 
\vspace{1.5cm}
{\large
{\sc A.\ Bredenstein$^1$, A.\ Denner$^2$, S.\ Dittmaier$^3$ 
and S.\ Pozzorini$^3$} } \\[.5cm]
$^1$ {\it High Energy Accelerator Research
                Organization (KEK),
\\
Tsukuba, Ibaraki 305-0801, Japan} \\[0.5cm]
$^2$ {\it Paul Scherrer Institut, W\"urenlingen und Villigen,
\\
CH-5232 Villigen PSI, Switzerland} \\[0.5cm]
$^3$ {\it Max-Planck-Institut f\"ur Physik
(Werner-Heisenberg-Institut), \\
D-80805 M\"unchen, Germany}
\par \vskip 1em
\end{center}\par
\vfill {\bf Abstract:} 
\par 
The process $\Pp\Pp\to\Pt\bar\Pt\Pb\bar\Pb+X$ represents a very
important background reaction to searches at the LHC, in particular to
$\Pt\bar\Pt\PH$ production where the Higgs boson decays into a
$\Pb\bar\Pb$ pair. A successful analysis of $\Pt\bar\Pt\PH$ at the LHC
requires the knowledge of direct $\Pt\bar\Pt\Pb\bar\Pb$ production at
next-to-leading order in QCD. We take the first step in this direction
upon calculating the next-to-leading-order QCD corrections to the
subprocess initiated by $q\bar q$ annihilation.  We devote an appendix
to the general issue of rational terms resulting from ultraviolet or
infrared (soft or collinear) singularities within dimensional
regularization.  There we show that, for arbitrary processes, in the
Feynman gauge, rational terms of infrared origin cancel in truncated
one-loop diagrams and result only from trivial self-energy
corrections.
\par
\vskip 2.5cm
\noindent
July 2008
\null
\setcounter{page}{0}
\clearpage
\def\thefootnote{\arabic{footnote}}
\setcounter{footnote}{0}

\section{Introduction}

The search for new particles will be the primary task of the LHC
experiment at CERN starting this year. The discovery of new particles
in the first place requires to establish excess of events over
background. The situation at the LHC is particularly complicated by
the fact that for many expected signals the corresponding background
cannot entirely be determined from data, but has to be assessed upon
combining measurements in signal-free regions with theory-driven
extrapolations. To this end, a precise prediction for the background
is necessary, in particular including next-to-leading-order (NLO)
corrections in QCD. Since many of these background processes involve
three, four, or even more particles in the final state, this kind of
background control requires NLO calculations at the technical
frontier. This problem lead to the creation of an ``experimenters'
wishlist for NLO calculations'' at the Les Houches workshop 2005
\cite{Buttar:2006zd}, updated in 2007 \cite{Bern:2008ef}, which
triggered great theoretical progress in recent years (see for instance
\citeres{Buttar:2006zd, Bern:2008ef, Ferroglia:2002mz, Denner:2002ii,
  Denner:2005nn, Giele:2004ub, Binoth:2005ff, Ossola:2006us,
  Lazopoulos:2007ix, Bern:2007dw, Ellis:2007br, Britto:2008vq,
  Berger:2008sj, Catani:2008xa, Giele:2008bc} and references therein).
Meanwhile the listed processes involving at most five particles in
loops have been completed in NLO QCD, including the production of
$\PW\PW+\mathrm{jet}$~\cite{Dittmaier:2007th,Campbell:2007ev},
weak-boson pairs plus two jets via vector-boson
fusion~\cite{Jager:2006zc}, and triple weak-boson
production~\cite{Lazopoulos:2007ix,Hankele:2007sb}.  However, none of
the true $2\to4$ processes has yet been addressed at NLO%
\footnote{Progress in the calculation of the virtual corrections to
  $\Pu\bar\Pu\to\Ps\bar\Ps\Pb\bar\Pb$ was reported in
  \citere{Binoth:2008gx}.}.  Among those processes,
$\Pp\Pp\to\Pt\bar\Pt\Pb\bar\Pb+X$ has top priority.  This process has
also been discussed as signal of strong electroweak symmetry breaking
\cite{Gintner:2008qs}.

The process of $\Pt\bar\Pt\Pb\bar\Pb$ production represents a very
important background to $\Pt\bar\Pt\PH$ production where the Higgs
boson decays into a $\Pb\bar\Pb$ pair.  While early studies of
$\Pt\bar\Pt\PH$ production at ATLAS~\cite{atlas-cms-tdrs} and
CMS~\cite{Drollinger:2001ym} suggested even discovery potential of
this process for a light Higgs boson, more recent
analyses~\cite{Cammin:2003,Cucciarelli:2006} with more realistic
background assessments show that the signal significance is
jeopardized if the background from $\Pt\bar\Pt\Pb\bar\Pb$ and
$\Pt\bar\Pt+\mathrm{jets}$ final states is not controlled very well.
This is a clear call for improved signal and background studies based
on NLO predictions to these complicated processes. For the
$\Pt\bar\Pt\PH$ signal~\cite{Beenakker:2001rj,Beenakker:2002nc} and
the $\Pt\bar\Pt+\mathrm{1jet}$ background~\cite{Dittmaier:2007wz} at
the LHC such predictions have been accomplished in recent years.

The dominant mechanism to produce $\Pt\bar\Pt\Pb\bar\Pb$ final states
in hadronic collisions is pure QCD. In leading order (LO)
quark--antiquark ($q\bar q)$ and gluon--gluon ($\Pg\Pg$) initial
states contribute, where the latter strongly dominate at the LHC
because of the high gluon flux.  Being of order
$\alpha^4_{\mathrm{s}}$ the corresponding cross sections are affected
by a very large scale uncertainty, which amounts to a factor two or
more.  Technically the $q\bar q$ channel is simpler to deal
with---though still demanding---and thus represents a natural first
step towards a full treatment of $\Pp\Pp\to\Pt\bar\Pt\Pb\bar\Pb+X$ at
NLO.  In this paper we report on this first step and present some
details of the calculation as well as numerical results.  These
results do not yet significantly improve the predictions for the LHC,
but on the one hand form a building block of the full calculation and
can serve as benchmark results for other groups on the other.
Moreover, this step proves the performance of the applied strategy and
methods, providing confidence that the more complicated $\Pg\Pg$
channel can be attacked widely in the same way.

In \refse{se:setup} we give a brief description of the NLO
calculation, followed by numerical results on integrated cross
sections in \refse{se:numres}.  Appendix~\ref{app:rat-terms} provides
a general discussion of rational terms in one-loop amplitudes, and
Appendix~\ref{app:SMEs} outlines some technical details concerning our
treatment of the Dirac algebra.  Finally, as a benchmark, in
Appendix~\ref{app:benchmark} we give numerical results for the matrix
element squared in lowest order and including virtual corrections for
one phase-space point.

\section{Details of the calculation}
\label{se:setup}

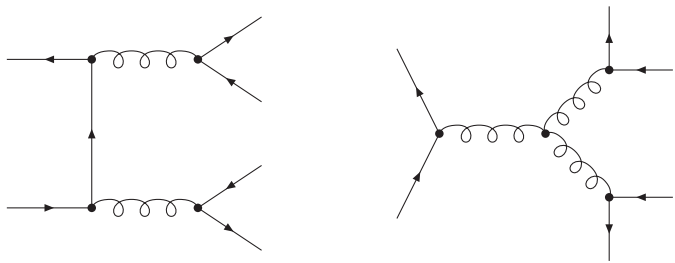
\begin{figure}[b]
\centerline{
{\setlength{\unitlength}{0.8pt}
\SetScale{0.8}
\begin{picture}(160,120) (-80,-60)
\ArrowLine(-60,-35)(-20,-35)
\Vertex(-20,-35){2}
\ArrowLine(-20,-35)(-20,35)
\Vertex(-20,35){2}
\ArrowLine(-20,35)(-60,35)
\Gluon(-20,35)(30,35){4}{3.5}
\Gluon(-20,-35)(30,-35){4}{3.5}
\ArrowLine(60,15)(30,35)
\Vertex(30,35){2}
\ArrowLine(30,35)(60,55)
\ArrowLine(60,-15)(30,-35)
\Vertex(30,-35){2}
\ArrowLine(30,-35)(60,-55)
\end{picture}
\hspace*{1em}
\begin{picture}(160,120) (-80,-60)
\ArrowLine(-60,-40)(-40,0)
\Vertex(-40,0){2}
\ArrowLine(-40,0)(-60,40)
\Vertex(10,0){2}
\Gluon(-40,0)(10,0){4}{3.5}
\Gluon(10,0)(40,30){4}{3.5}
\Gluon(10,0)(40,-30){4}{3.5}
\ArrowLine(70,30)(40,30)
\Vertex(40,30){2}
\ArrowLine(40,30)(40,60)
\ArrowLine(70,-30)(40,-30)
\Vertex(40,-30){2}
\ArrowLine(40,-30)(40,-60)
\end{picture}
}}
\caption{Two different diagram topologies contributing to
$q\bar q\to\Pt\bar\Pt\Pb\bar\Pb$ in LO QCD; there are six explicit diagrams
of the first and one of the second kind.}
\label{fig:LOtops}
\end{figure}

\begin{figure}
\centerline{
{\setlength{\unitlength}{0.8pt}
\SetScale{0.8}
\begin{picture}(170,130)(-90,-65)
\ArrowLine(-95,-40)(-75,0)
\Vertex(-75,0){2}
\ArrowLine(-75,0)(-95,40)
\Gluon(-75,0)(-40,0){4}{3.5}
\Vertex(-40,0){2}
\Gluon(-40,0)(0,45){4}{3.5}
\Gluon(-40,0)(0,-45){4}{3.5}
\Gluon(40,-25)(40,25){4}{3.5}
\ArrowLine(70,45)(40,25)
\Vertex(40,25){2}
\ArrowLine(40,25)(0,45)
\Vertex(0,45){2}
\ArrowLine(0,45)(30,65)
\ArrowLine(70,-45)(40,-25)
\Vertex(40,-25){2}
\ArrowLine(40,-25)(0,-45)
\Vertex(0,-45){2}
\ArrowLine(0,-45)(30,-65)
\end{picture}
\hspace*{1em}
\begin{picture}(160,130) (-80,-65)
\ArrowLine(-40,-25)(-40,25)
\Vertex(-40,25){2}
\ArrowLine(-40,25)(-70,25)
\Vertex(-40,-25){2}
\ArrowLine(-70,-25)(-40,-25)
\Gluon(-40,25)(0,50){4}{3.5}
\Gluon(-40,-25)(0,-50){4}{3.5}
\Gluon(40,-25)(40,25){4}{3.5}
\ArrowLine(70,45)(40,25)
\Vertex(40,25){2}
\ArrowLine(40,25)(0,50)
\Vertex(0,50){2}
\ArrowLine(0,50)(30,70)
\ArrowLine(70,-45)(40,-25)
\Vertex(40,-25){2}
\ArrowLine(40,-25)(0,-50)
\Vertex(0,-50){2}
\ArrowLine(0,-50)(30,-70)
\end{picture}}}
\caption{Diagram topologies for pentagon and hexagon graphs
contributing to $q\bar q\to\Pt\bar\Pt\Pb\bar\Pb$ at one loop in QCD; 
there are 24 explicit pentagons and 8 hexagons.}
\label{fig:NLOtops}
\end{figure}
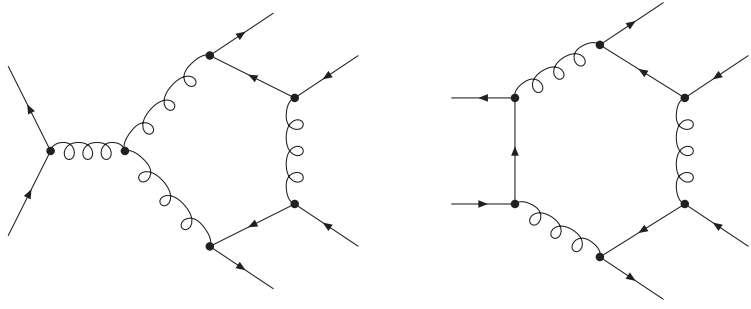

In LO QCD seven different Feynman diagrams contribute to the 
partonic process
$q\bar q\to\Pt\bar\Pt\Pb\bar\Pb$; the various topologies are shown
in \reffi{fig:LOtops}. 
The virtual QCD corrections comprise about 200 one-loop diagrams, the
most complicated being the 8~hexagons and 24~pentagons, which are
illustrated in \reffi{fig:NLOtops}.
The real QCD corrections in the $q\bar q$ channel are induced by gluon
bremsstrahlung, $q\bar q\to\Pt\bar\Pt\Pb\bar\Pb \Pg$, where the
corresponding 64~diagrams are obtained from the LO graphs upon adding
an external gluon in all possible ways.  In the following we briefly
describe the calculation of the virtual and real corrections, where
each of these contributions has been worked out twice and
independently, resulting in two completely independent computer codes.

\subsection{Virtual corrections}

The general strategy for the evaluation of the one-loop corrections is
based on the reduction of the amplitude $\M^{(\Ga)}$ of each
(sub)diagram $\Ga$ in the following way,
\beq\label{colspinreduction}
\M^{(\Ga)} = 
 {\cal C}^{(\Ga)}
\left(\sum_m {\cal F}^{(\Ga)}_m(\{\papb\}) \, \hat\M_m(\{p_a\})\right),
\eeq
where the colour structure ${\cal C}^{(\Ga)}$ present in the
(sub)diagram is factorized from the remaining colour-independent part.
The decomposition of the colour structure,
\beq
 {\cal C}^{(\Ga)}=
\sum_{k=1}^6 c^{(\Ga)}_k {\cal C}_k
,
\eeq
is done in a basis $\{{\cal C}_k\}$  consisting of six elements,
which can be chosen as
\beqar\label{colourbasis}
&
{\cal C}_1= 
1\otimes T^a \otimes T^a
,\qquad
{\cal C}_2= 
T^a\otimes 1 \otimes T^a
,\qquad
{\cal C}_3= 
T^a\otimes T^a \otimes 1
,&\nl&
{\cal C}_4= 
1\otimes 1 \otimes 1
,\qquad
{\cal C}_5= 
f^{abc}\, T^a\otimes T^b \otimes T^c
,\qquad
{\cal C}_6= 
d^{abc}\,T^a\otimes T^b \otimes T^c.&
\eeqar
Here $T^a$, $f^{abc}$, $d^{abc}$ are the usual SU(3) objects, and the
tensor products connect the three fermionic chains.  The LO amplitude
is decomposed in colour space as
\beq
\M^{\LO} = \sum_{l=1}^6  {\cal C}_l \,
{\cal M}^{\LO}_{l}.
\eeq
In general each loop diagram gives rise to $3^{n_4}$ colour-factorized
amplitudes of type \refeq{colspinreduction}, where $n_4$ is the number
of quartic gluon vertices in the diagram. However, for most diagrams
$n_4=0$, and the colour structure factorizes completely.  The colour
separation implies that the computation time for individual loop
diagrams does not scale with the number of colour structures present
in the basis $\{{\cal C}_k\}$.

The colour-free parts of $\M^{(\Ga)}$ are written as a linear
combination of so-called standard matrix elements (SMEs)
$\hat\M_m(\{p_a\})$, which contain all Dirac chains and the
polarization information.  Since the computing time scales with the
number of SMEs, it is important to reduce the set of SMEs
$\{\hat\M_m\}$ as much as possible.  To this end, we employ an
algebraic procedure based on four-dimensional relations that are
derived from Chisholm's identity whenever their use is admitted.  For
massless external fermions this four-dimensional reduction has been
described in detail in Sections~3.1 and 3.3 of \citere{Denner:2005fg};
here we had to generalize this approach to one massive and two
massless spinor chains. In this way some thousand different spinor
chains are reduced to about 150 SMEs $\hat\M_m(\{p_a\})$.  A brief
description of this procedure, which is implemented in two independent
{\sc Mathematica} programs, is outlined in \refapp{app:SMEs}.

The one-loop correction to the spin- and colour-summed squared
amplitude induced by $\Ga$ reads
\beq
2\Re\,\Biggl\{ 
\sum_{\mathrm{col}}\sum_{\mathrm{pol}}
\M^{(\Ga)} \left(\M^{\LO}\right)^*\Biggr\} = 
2\Re\,\left\{  
\sum_{k=1}^6 c^{(\Ga)}_k 
\sum_m {\cal F}^{(\Ga)}_m(\{\papb\}) \, M_{km}(\{\papb\})\right\},
\eeq
where the interference of the  LO amplitude with the elements of the 
SME and colour basis,
\beqar
M_{km}(\{\papb\}) 
&=& 
\sum_{\mathrm{col}}
 {\cal C}_k
\sum_{\mathrm{pol}}
\hat\M_m(\{p_a\}) \left(\M^{\LO}\right)^*
\nl
&=& 
\sum_l
\sum_{\mathrm{col}}
 {\cal C}_k
 {\cal C}^*_l
\sum_{\mathrm{pol}}
\hat\M_m(\{p_a\}) \left(\M_l^{\LO}\right)^*,
\eeqar
has to be calculated only once per phase-space point.  Moreover, the
colour-correlation matrix $\sum_{\mathrm{col}} {\cal C}_k {\cal
  C}^*_l$ obviously is independent of the kinematics and is only
calculated once and for all.
The most time-consuming components of the numerical calculation are
the scalar form factors ${\cal F}^{(\Ga)}_m$, which are linear
combinations of the Lorentz-invariant coefficients of $N$-point tensor
loop integrals with rank $R \le 3$ and degree $N\le 6$,
\beq
{\cal F}^{(\Ga)}_m(\{\papb\}) = 
\sum_{R} \,
\sum_{j_1,\dots,j_R}
{\cal K}^{(\Ga)}_{m;j_1,\dots,j_R}(\{\papb\})
\,T^{N}_{j_1,\dots,j_R}(\{\papb\})
.
\eeq
The evaluation of one-loop tensor integrals  $T^{N}_{j_1,\dots,j_R}$
follows the strategy of \citeres{Denner:2002ii,Denner:2005nn}%
\footnote{We note in passing that the reduction methods of
  \citeres{Denner:2002ii,Denner:2005nn} have also been used in the
  related calculation~\cite{Lei:2007rv} of NLO QCD corrections to the
  $2\to4$ particle process $\ga\ga\to\Pt\bar\Pt\Pb\bar\Pb$ at a
  $\ga\ga$ collider.}  that was already successfully used to compute
the NLO electroweak corrections to $\Pep\Pem\to4\,$fermions
\cite{Denner:2005fg,Denner:2005es}.  In this approach the analytic
expressions are not reduced to master integrals.  In contrast, the
tensor integrals are evaluated by means of algorithms that perform a
recursive reduction to master integrals in numerical form.  This
avoids huge analytic expressions and permits to adapt the reduction
strategy to the specific numerical problems that appear in different
phase-space regions.

\begin{sloppypar}
The scalar master integrals are evaluated using the methods and
results of \citeres{'tHooft:1979xw,Beenakker:1990jr}.  Ultraviolet
(UV) divergences are regularized
dimensionally in both evaluations, but infrared (IR) divergences are
treated in different ways as described below.  Following ideas from
the 1960's~\cite{Me65}, tensor and scalar 6-/5-point functions are
directly expressed in terms of 5-/4-point integrals
\cite{Denner:2002ii,Denner:2005nn}.%
\footnote{Similar reductions are described in \citere{Binoth:2005ff}.}
Tensor 4-point and 3-point integrals are reduced to scalar integrals
with the Passarino--Veltman algorithm \cite{Passarino:1979jh} as long
as no small Gram determinant appears in the reduction.  If small Gram
determinants occur, two alternative
schemes are applied \cite{Denner:2005nn}.%
\footnote{Similar procedures based on numerical evaluations of
  specific one-loop integrals~\cite{Binoth:2005ff,Ferroglia:2002mz} or
  expansions in small determinants~\cite{Giele:2004ub} have also been
  proposed by other authors.}  One method makes use of expansions of
the tensor coefficients about the limit of vanishing Gram determinants
and possibly other kinematical determinants.  In the second
(alternative) method we evaluate a specific tensor coefficient, the
integrand of which is logarithmic in Feynman parametrization, by
numerical integration. Then the remaining coefficients as well as the
standard scalar integral are algebraically derived from this
coefficient.  The results of the two different codes, based on the
different methods described above are in good numerical agreement.
Although both versions of the virtual corrections basically follow the
same strategy for the evaluation of loop integrals, they are based on
independent in-house libraries.  In each of the two calculations the
cancellation of IR and UV singularities was checked with high
precision in the numerical results.
\end{sloppypar}

{\it Version~1} of the virtual corrections starts with the generation
of Feynman diagrams using {\sc FeynArts}~1.0~\cite{Kublbeck:1990xc}.
Their algebraic reduction is completely performed with in-house {\sc
  Mathematica} routines. In detail, $D$-dimensional identities (Dirac
algebra, Dirac equation) are used until UV divergences cancel against
counterterms. IR (soft and collinear) divergences are regularized
dimensionally and separated from full diagrams in terms of 3-point
subdiagrams as described in \citere{Dittmaier:2003bc}.  We subtract
the IR-divergent part $\M^{(\Ga,D)}_{\mathrm{sing}}$ from the
amplitude $\M^{(\Ga,D)}$ of a \mbox{(sub-)}diagram $\Ga$, where $D$
indicates dimensional regularization, and add it back.  Note that
$\M^{(\Ga,D)}_{\mathrm{sing}}$ can be easily constructed already at
the integrand level of the whole diagram following
\citere{Dittmaier:2003bc}.  In the IR-finite and
regularization-scheme-independent difference
$\M^{(\Ga,D)}-\M^{(\Ga,D)}_{\mathrm{sing}}$ we can then switch from
dimensional regularization to a four-dimensional scheme,
\beq
\M^{(\Ga,D)} = 
\left(\M^{(\Ga,D)}-\M^{(\Ga,D)}_{\mathrm{sing}}\right)
+\M^{(\Ga,D)}_{\mathrm{sing}}
= \left(\M^{(\Ga,\la)}-\M^{(\Ga,\la)}_{\mathrm{sing}}\right)
+\M^{(\Ga,D)}_{\mathrm{sing}},
\eeq
where $\la$ indicates any mass regulators in four dimensions.
Specifically, we introduce infinitesimal light-quark and gluon masses
with the hierarchy $\Mg\ll m_q$ to regularize the IR singularities.
The evaluation of $\M^{(\Ga,\la)}$ then proceeds in $D=4-2\eps$
dimension merely to regularize the UV singularities, i.e.\ so-called
rational terms resulting from $(D-4)$ times poles in $\eps$ have to be
taken care of for UV singularities, but not for IR singularities in
this part. Possible rational terms of IR origin are contained in
$\M^{(\Ga,D)}_{\mathrm{sing}}$ which entirely consists of 3-point
subgraphs and is thus easy to reduce to scalar 2- and 3-point
integrals $B_0$ and $C_0$, thereby keeping the full dependence on $D$.
It turns out that no $D$-dependent prefactors occur in front of
IR-singular integrals. In \refapp{app:rat-terms} we show that this
result of our specific calculation is not accidental, but generalizes
to arbitrary processes at NLO.  Technically it is easier to evaluate
$\M^{(\Ga,D)}_{\mathrm{sing}}$ and $\M^{(\Ga,\la)}_{\mathrm{sing}}$
simultaneously according to
\beq
\M^{(\Ga,D)}_{\mathrm{sing}}-\M^{(\Ga,\la)}_{\mathrm{sing}}
= \M^{(\Ga,D)}_{\mathrm{sing}}
\Big|_{B_0^{(D)}\to\De B_0,C_0^{(D)}\to\De C_0}
\equiv \De\M^{(\Ga)}_{\mathrm{sing}},
\eeq
where $\De I=I^{(D)}-I^{(\la)}$ are the differences of the scalar
integrals $I=B_0,C_0$ in the two IR regularization schemes.  Note that
IR-finite integrals drop out in $\De\M^{(\Ga)}_{\mathrm{sing}}$
completely.  Having cancelled UV divergences against counterterms and
controlled the $D$-dimensional issues concerning IR singularities, the
amplitude is further simplified in four space--time dimensions.
Specifically, the reduction of SMEs described in \refapp{app:SMEs} is
performed then.  The IR-divergent ``endpoint part'' of the dipole
subtraction function, i.e.\ the contribution of the $I$ operator as
defined in \citere{Catani:2002hc}, is processed through the described
algebraic manipulations in the same way as LO and one-loop amplitudes.
The algebraic {\sc Mathematica} output of each diagram is
automatically processed to {\sc Fortran} for the numerical evaluation.

{\it Version~2} of the virtual corrections employs {\sc
  FeynArts}~3.2~\cite{Hahn:2000kx} for generating and {\sc
  FormCalc}~5.2~\cite{Hahn:1998yk} for preprocessing the amplitudes.
The first part of the calculation is performed in $D$ dimensions.  In
particular, the so-called rational terms resulting from the UV
divergences of tensor loop coefficients are automatically extracted by
{\sc FormCalc}.  Since the IR divergences that appear in the $q\bar q$
channel are of abelian nature, we exploit the fact that they can be
regularized as in QED by means of fermion and gauge-boson (gluon)
masses, $m_q$ and $\Mg$.  These masses are treated as infinitesimal
quantities (with $\Mg\ll m_q$) both in the algebraic expressions and
in the numerical routines that evaluate the tensor integrals, \ie only
the logarithmic dependence on these mass parameters is retained.
Corresponding IR singularities associated with real emission have been
obtained from \citere{Catani:2002hc} by means of an appropriate change
of regularization scheme.

Being of diagrammatic nature, the employed techniques are sometimes
denoted as ``brute force" methods.  This choice of the terminology
might suggest scarce efficiency.  In fact, the performance of the
algorithms is a very important issue that should be assessed by means
of those quantities that describe the problematic aspects of NLO
multi-leg calculations: numerical accuracy and CPU time.  In this
respect our treatment of the virtual corrections is characterized by
high numerical precision and speed.  The numerical agreement between
the two programs is good, and the CPU time needed to evaluate a
phase-space point (including sums over colours and polarizations)
amounts to about $10^{-2}$ seconds on a single 3\,GHz Intel Xeon
processor.  This provides a benchmark that can be compared with the
efficiency of other approaches.

\subsection{Real corrections}

In both evaluations of the real corrections the amplitudes are
calculated in the form of helicity matrix elements.  The singularities
for soft or collinear gluon emission are isolated via dipole
subtraction~\cite{Catani:2002hc,Catani:1996vz,Dittmaier:1999mb,Phaf:2001gc}
for NLO QCD calculations using the formulation~\cite{Catani:2002hc}
for massive quarks.  After combining virtual and real corrections,
singularities connected to collinear configurations in the final state
cancel for ``collinear-safe'' observables automatically after applying
a jet algorithm, singularities connected to collinear initial-state
splittings are removed via $\overline{\mathrm{MS}}$ QCD factorization
by PDF redefinitions.  While soft and collinear singularities have to
be regularized in the ``endpoint part'' of the subtraction function,
\ie the part of the subtraction terms that has to be combined with the
virtual corrections, no regularization is needed in the subtraction
terms for the real corrections.  In both evaluations the phase-space
integration is performed with multichannel Monte Carlo generators
\cite{Berends:1994pv} and adaptive weight optimization similar to the
one implemented in {\sc RacoonWW} \cite{Denner:1999gp}.

In {\it version~1} of the real corrections the matrix elements have
been calculated using the Weyl--van-der-Waerden spinor technique in
the formulation of \citere{Dittmaier:1998nn}. Soft and collinear
singularities are regularized using dimensional regularization.  The
phase-space integration, implemented in {\sc C++}, is based on {\sc
  RacoonWW}, but the phase-space mappings are built up in a more
generic way very similar to the approach of {\sc
  Lusifer}~\cite{Dittmaier:2002ap}.

In {\it version~2} of the real corrections the matrix elements have
been generated with {\sc Madgraph 4.1.33} \cite{Stelzer:1994ta}. As in
the corresponding virtual corrections, soft singularities are
regularized by an infinitesimal gluon mass and collinear singularities
by small quark masses, which appear only in logarithms in the endpoint
part of the subtraction function.  The Monte Carlo generator is a
further development of the one used in {\sc COFFER$\ga\ga$}
\cite{Bredenstein:2005zk} and for the calculation of the NLO
corrections to $\Pp\Pp\to\PH+\mathrm{2jets}+X$
\cite{Ciccolini:2007jr}.

In version~2 we have also implemented two-cut-off slicing for the
purpose of checking. In this approach (as \eg reviewed in
\citere{Harris:2001sx}), phase-space regions where real gluon emission
contains soft or collinear singularities are defined by the auxiliary
cutoff parameters $\delta_\mathrm{s} ,\; \delta_\mathrm{c}\ll 1$ in
the partonic centre-of-mass frame. In real gluon radiation processes,
the region
\beq
 m_{\mathrm{g}} < k^0 < \delta_\mathrm{s} \frac{\sqrt{\hat s}}{2},
\eeq 
where $k$ is the gluon momentum and $\sqrt{\hat s}$ the partonic
centre-of-mass energy,  
is treated in soft approximation.  The regions determined by 
\beq \label{eq:coll} 
1-\cos(\theta_{\Pg q}) < \delta_\mathrm{c},
 \qquad k^0 > \delta_\mathrm{s} \frac{\sqrt{\hat s}}{2},
 \label{eq:slicingcuts} 
\eeq 
where $\theta_{\Pg q}$ is the angle between any light quark $q$
(including b quarks) and the gluon, are evaluated using collinear
factorization. We again use an infinitesimal gluon mass and quark
masses as regulators. In this regularization the contributions of the
soft regions for light quarks and of the collinear regions can be
found in \citere{Denner:2000bj}, and the contributions of the soft
regions involving top quarks can easily be calculated with the
explicit results for the soft integrals in
\citeres{'tHooft:1979xw,Denner:1991kt}. In the remaining phase space
no regulators are used.  When adding all contributions, the dependence
on the technical cuts cancels if the cut-off parameters are chosen to
be small enough so that the soft and collinear approximations apply,
\ie the slicing result is correct up to terms of ${\cal
  O}(\delta_\mathrm{s})$ and ${\cal O}(\delta_\mathrm{c})$. Since the
numerical cancellations between the different contributions grow with
smaller cut parameters, the numerical error blows up if these
parameters are too small.

\section{Numerical results}
\label{se:numres}

We consider the process $\Pp\Pp\to\Pt\bar\Pt\Pb\bar\Pb+X$ at the LHC,
\ie for $\sqrt{s}=14\TeV$.  For the top-quark mass, renormalized in
the on-shell scheme, we take the numerical value $\Mt=172.6\GeV$
\cite{Group:2008nq}. All other QCD partons (including $\Pb$~quarks)
are treated as massless particles, and collinear final-state
configurations, which give rise to singularities, are recombined into
IR-safe jets using a \mbox{$k_{\rT}$-algorithm} \cite{Catani:1992zp}.
Specifically, we adopt the \mbox{$k_\rT$-algorithm} of
\citere{Blazey:2000qt} and recombine all final-state b quarks and
gluons with pseudorapidity $|\eta| < 5$ into jets with separation
$\sqrt{\Delta\phi^2+\Delta y^2}>D=0.8$ in the
rapidity--azimuthal-angle plane.  Requiring two b-quark jets, this
also avoids collinear singularities resulting from the splitting of
gluons into (massless) b quarks.  Motivated by the search for a
$\Pt\bar \Pt H (H\to\Pb\bar \Pb)$ signal at the LHC
\cite{Cammin:2003,Cucciarelli:2006}, we impose the following
additional cuts on the transverse momenta, the rapidity, and the
invariant mass of the two (recombined) b-jets:%
\footnote{The experimental analysis of $\Pt\bar \Pt H (H\to\Pb\bar
  \Pb)$ will select b quarks with transverse momenta much higher than
  $\Mb$, justifying the approximation $\Mb=0$.}
\mbox{$p_{\rT,\Pb}>20\GeV$}, \mbox{$|y_\Pb|<2.5$}, and
$m_{\Pb\bar\Pb}> m_{\Pb\bar\Pb,\mathrm{cut}}$. We plot results either
as a function of $m_{\Pb\bar\Pb,\mathrm{cut}}$ or for
$m_{\Pb\bar\Pb,\mathrm{cut}}=0$.  Note, however, that the jet
algorithm and the requirement of having two b~jets with 
$p_{\rT,\Pb}>20\GeV$ in the final state sets an effective lower limit
on the invariant mass $m_{\Pb\bar\Pb}$ of roughly $20\GeV$.  The
outgoing (anti)top quarks are neither affected by the jet algorithm
nor by phase-space cuts.

We consistently use the CTEQ6 \cite{Pumplin:2002vw} set of parton
distribution functions (PDFs), i.e.\ we take CTEQ6L1 PDFs with a
1-loop running $\alpha_{\mathrm{s}}$ in LO and CTEQ6M PDFs with a
2-loop running $\alpha_{\mathrm{s}}$ in NLO, but the suppressed
contribution from b~quarks in the initial state has been neglected.
The number of active flavours is $N_{\mathrm{F}}=5$, and the
respective QCD parameters are $\Lambda_5^{\mathrm{LO}}=165\MeV$ and
$\Lambda_5^{\overline{\mathrm{MS}}}=226\MeV$.  In the renormalization
of the strong coupling constant the top-quark loop in the gluon
self-energy is subtracted at zero momentum. In this scheme the running
of $\alpha_{\mathrm{s}}$ is generated solely by the contributions of
the light-quark and gluon loops.  This yields
$\alpha_{\mathrm{s}}(\Mt)|_{\mathrm{LO}} =0.1178730\ldots$ and
$\alpha_{\mathrm{s}}(\Mt)|_{\mathrm{NLO}}=0.1076396\ldots$.  By
default, we set the renormalization and factorization scales,
$\mu_{\mathrm{R}}$ and $\mu_{\mathrm{F}}$, to the common value
$\mu_0=\Mt+m_{\Pb\bar\Pb,\mathrm{cut}}/2$.

\subsection{Integrated cross sections}

\begin{figure}
\centerline{
\includegraphics[bb= 85 440 285 660, width=.5\textwidth]
{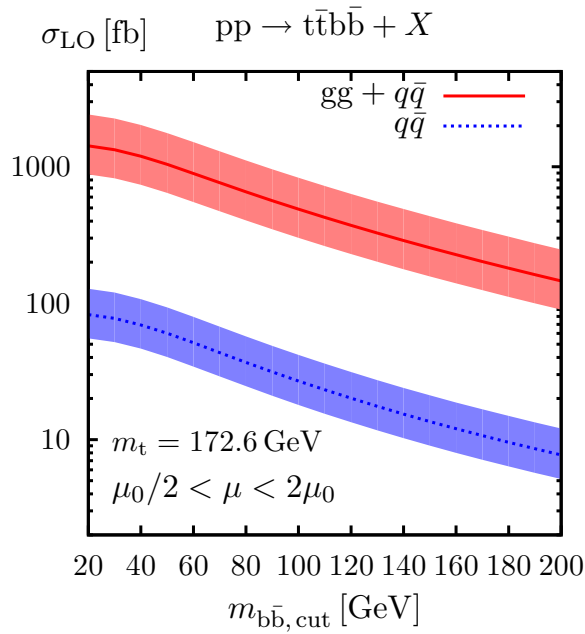}
}
\caption{LO cross section ($\Pg\Pg+q\bar q$) versus contribution from
  $q\bar q$ annihilation for $\Pp\Pp\to\Pt\bar\Pt\Pb\bar\Pb+X$ at the
  LHC as function of the cut $m_{\Pb\bar\Pb,\mathrm{cut}}$ on the
  invariant mass of the $\Pb\bar\Pb$ pair.}
\label{fig:LOcsmbb}
\end{figure}

We first consider results for integrated cross sections.
Figure~\ref{fig:LOcsmbb} shows the total LO cross section and the
contribution induced by $q\bar q$ annihilation at the LHC as a
function of $m_{\Pb\bar\Pb,\mathrm{cut}}$.
For the chosen setup $\Pg\Pg$ fusion dominates over the $q\bar q$
channel by roughly a factor 17.  The renormalization and factorization
scale dependence of the LO prediction is indicated by bands resulting
from varying the central scale $\mu_0$ up and down by a factor 2
which corresponds to a variation of the cross section by a factor 1.6.
Owing to the large power of $\alpha_{\mathrm{s}}(\mu_{\mathrm{R}})^4$
in the LO cross section the scale uncertainty is strongly dominated by
the renormalization scale dependence.  We note that the LO cross
sections have also been reproduced with the program {\sc
  Sherpa}~\cite{Gleisberg:2003xi}.

\begin{figure}
\centerline{
\includegraphics[bb= 85 440 285 660, width=.5\textwidth]{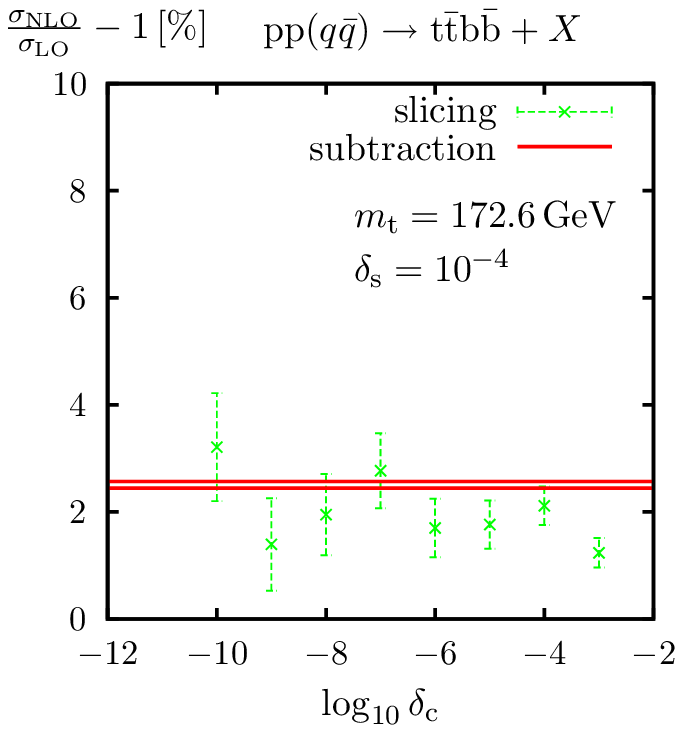}
\includegraphics[bb= 85 440 285 660, width=.5\textwidth]{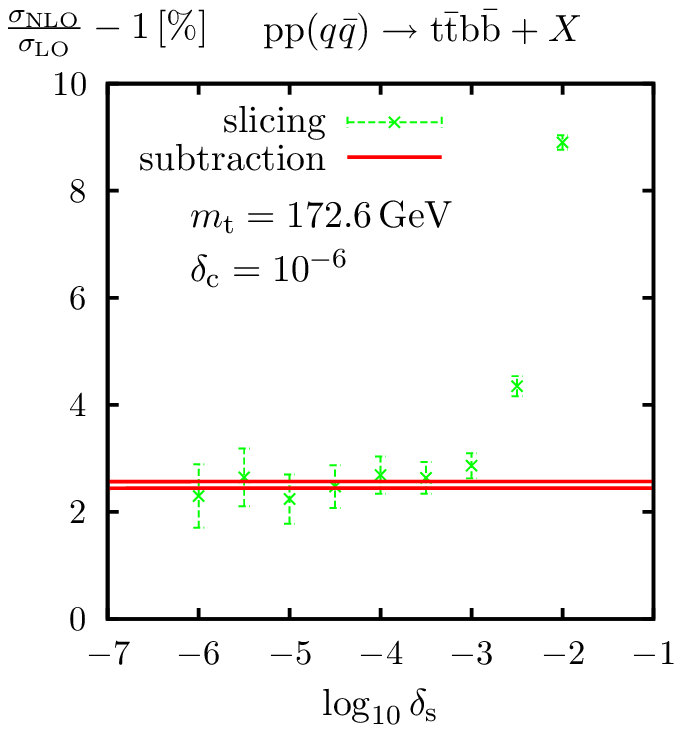}
}
\caption{Comparison of the relative NLO corrections to
$\Pp\Pp(q\bar q)\to\Pt\bar\Pt\Pb\bar\Pb+X$ at the LHC as obtained with
dipole subtraction and two-cutoff phase-space slicing
using $m_{\Pb\bar\Pb,\mathrm{cut}}=0$ and
$\mu_{\mathrm{R}}=\mu_{\mathrm{F}}=\mu_0=\Mt$.}
\label{fig:subsli}
\end{figure}

Figure~\ref{fig:subsli} illustrates the mutual agreement between 
NLO results obtained with dipole subtraction and two-cutoff 
phase-space slicing.
We find that within integration errors the slicing results become
independent of the cut-offs for $\delta_\mathrm{s}\lsim10^{-3}$ and
$\delta_\mathrm{c}\lsim10^{-4}$ and agree nicely with the result of
the subtraction method.  While the latter has been obtained with
$2\times10^8$ events, the slicing results are based on $10^9$ events.
Still the statistical error obtained with the subtraction approach
(indicated by the width of the band) is almost an order of magnitude
smaller than its slicing counterpart (errorbars), demonstrating the
higher efficiency of dipole subtraction.  The results shown in the
following are obtained with the subtraction approach.

\begin{figure}
\centerline{
\includegraphics[bb= 85 440 285 660, width=.5\textwidth]{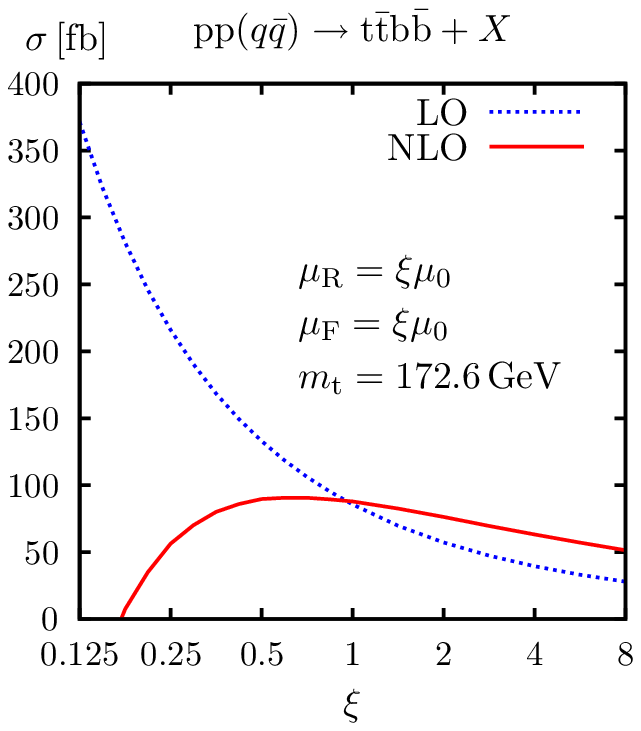}
\includegraphics[bb= 85 440 285 660, width=.5\textwidth]{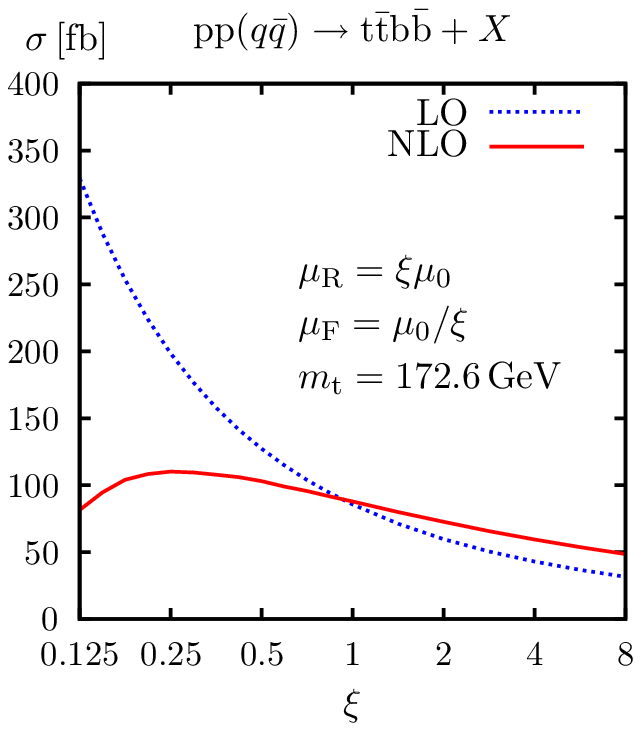}
}
\caption{Dependence of the LO and NLO cross sections of
$\Pp\Pp(q\bar q)\to\Pt\bar\Pt\Pb\bar\Pb+X$ at the LHC
for $m_{\Pb\bar\Pb,\mathrm{cut}}=0$ and $\mu_0=\Mt$.}
\label{fig:scaledep}
\end{figure}

In \reffi{fig:scaledep} we show the scale dependence of the LO and NLO
cross sections induced by the $q\bar q$ channel upon varying the
renormalization and factorization scales in a uniform or an antipodal
way.
We observe a sizeable reduction of the scale uncertainty upon going
from LO to NLO. Varying the scale up and down by a factor 2 changes
the cross section by 55\% in LO and by 17\% in NLO. At the central
scale, the NLO correction is small, \ie $\sim2.5\%$, and the LO and
NLO cross sections are given by $\si_{\mathrm{LO}}=85.522(26)\fb$ and
$\si_{\mathrm{NLO}}=87.698(56)\fb$.  The numbers in parentheses are
the errors of the Monte Carlo integration for $2\times10^8$ events,
where the virtual corrections are only calculated for each 5th event.

\begin{figure}
\centerline{
\includegraphics[bb= 85 440 285 660, width=.5\textwidth]
{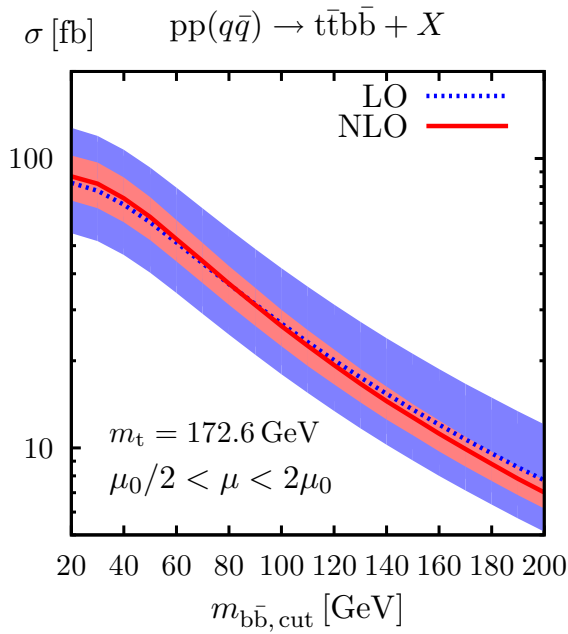}
}
\caption{LO and NLO cross sections 
  for $\Pp\Pp(q\bar q)\to\Pt\bar\Pt\Pb\bar\Pb+X$ at the LHC as
  function of the cut $m_{\Pb\bar\Pb,\mathrm{cut}}$ on the invariant
  mass of the $\Pb\bar\Pb$ pair, with the bands indicating the scale
  dependence by varying $\mu_{\mathrm{R}}$ and $\mu_{\mathrm{F}}$ by
  factors $1/2$ and $2$ in a uniform or antipodal way.}
\label{fig:NLOcsmbb}
\end{figure}

Figure~\ref{fig:NLOcsmbb} shows the LO and NLO cross sections as
function of the cut $m_{\Pb\bar\Pb,\mathrm{cut}}$ on the $\Pb\bar\Pb$
invariant mass, where the bands indicate the effect from a uniform or
antipodal rescaling of $\mu_{\mathrm{R}}$ and $\mu_{\mathrm{F}}$ by
factors $1/2$ and $2$.
The reduction of the scale uncertainty from about $\pm50\%$ to
$\pm17\%$ and the smallness of the NLO correction holds true for the
considered range in $m_{\Pb\bar\Pb,\mathrm{cut}}$, which is motivated
by the search for a low-mass Higgs boson.  While the NLO prediction is
consistent with the LO uncertainty band, the shape of the distribution
is distorted by the corrections. For the central scale we find an NLO
correction of $+2.5\%$ for small $m_{\Pb\bar\Pb,\mathrm{cut}}$ but a
correction of $-11\%$ for $m_{\Pb\bar\Pb,\mathrm{cut}}=200\GeV$.

\subsection{Differential cross sections}

In this section we consider results for distributions in variables
related to the $\Pb\Pbbar$ pair (which in the corresponding signal
process $\Pp\Pp\to\Pt\Ptbar\PH+X$ results from the Higgs decay).  For
each distribution we plot the absolute predictions in LO and in NLO
and show the relative corrections. These results are based on
$2\times10^8$ events, and no cut on $m_{\Pb\bar\Pb}$ has been applied
such that the default scale is $\mu_0=\Mt$.

\begin{figure}
\includegraphics[bb= 85 440 285 660, width=.5\textwidth]{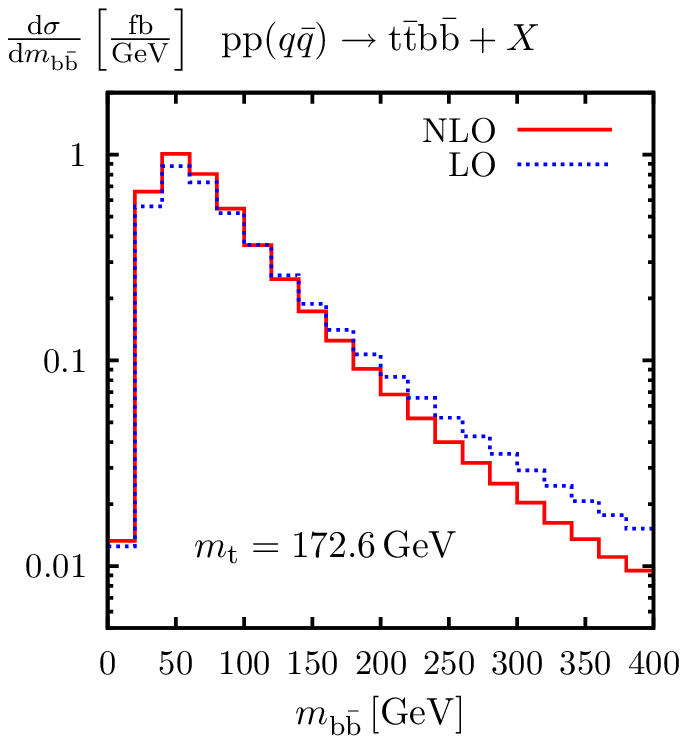}%
\includegraphics[bb= 85 440 285 660, width=.5\textwidth]{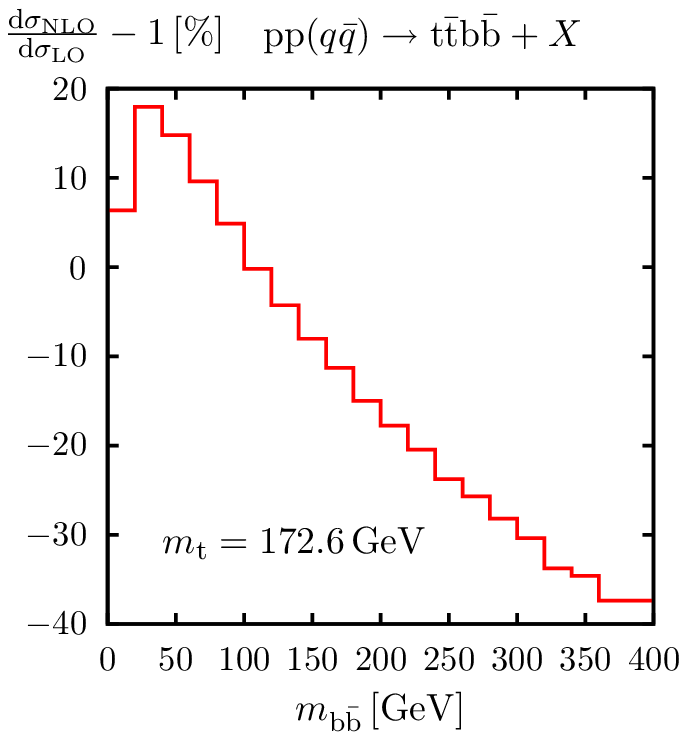}
\caption{Distribution in the invariant mass $m_{\Pb\Pbbar}$ of the
  bottom--antibottom pair (left) and corresponding relative NLO
  corrections (right) for $\mu_{\mathrm{R}}=\mu_{\mathrm{F}}=\mu_0=\Mt$.}
\label{fi:distr-mbb}
\end{figure}

We first show the distribution in the invariant mass $m_{\Pb\bar\Pb}$
of the $\Pb\Pbbar$ pair in \reffi{fi:distr-mbb}.
The differential cross section drops strongly with increasing
$m_{\Pb\bar\Pb}$, while the relative NLO corrections become large and
negative. The increase of the corrections with $m_{\Pb\bar\Pb}$ is
larger than the one seen in \reffi{fig:NLOcsmbb} since the scales are
fixed to $\Mt$ and not related to $m_{\Pb\bar\Pb}$.  In this paper we
do not investigate in how far shape distortions induced by the
corrections could be absorbed into the LO upon using
phase-space-dependent scales; we postpone this issue until the full
NLO corrections including $\Pg\Pg$ fusion are available.  The drop of
the distribution for small $m_{\Pb\bar\Pb}$ is due to the fact that
the jet algorithm provides an effective cut on this variable.

\begin{figure}
\includegraphics[bb= 85 440 285 660, width=.5\textwidth]{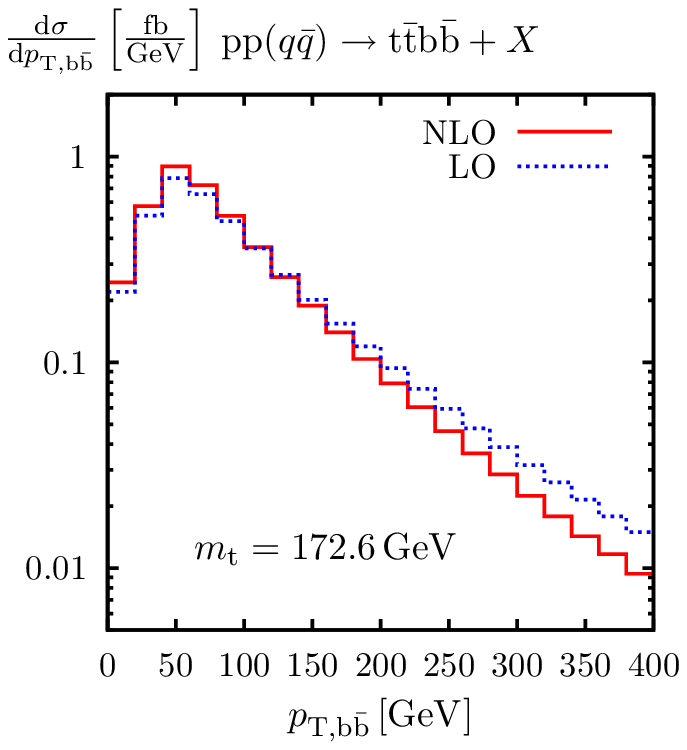}%
\includegraphics[bb= 85 440 285 660, width=.5\textwidth]{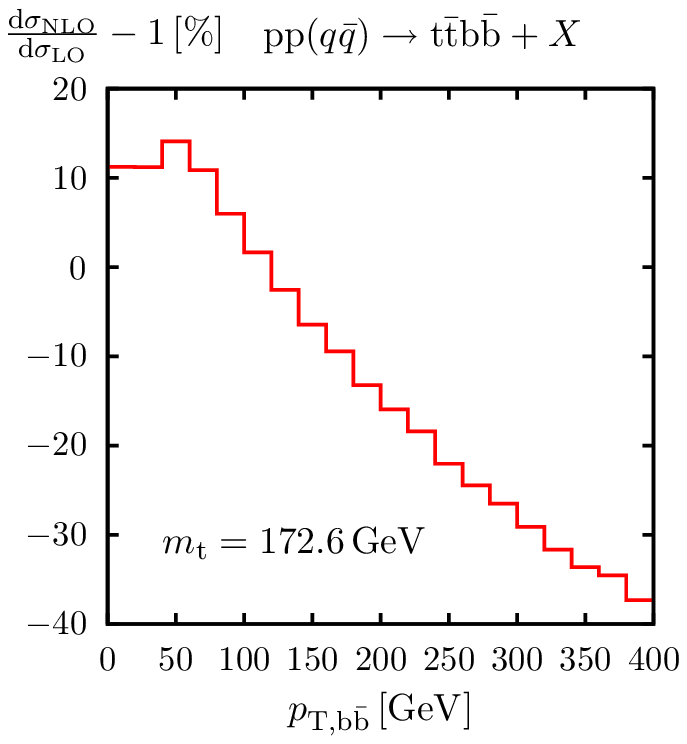}
\caption{Distribution in the transverse momentum $p_{\rT,\Pb\Pbbar}$ of the
  bottom--antibottom pair (left) and corresponding relative corrections
  (right) for $\mu_{\mathrm{R}}=\mu_{\mathrm{F}}=\mu_0=\Mt$.}
\label{fi:distr-ptbb}
\end{figure}

The distribution in the transverse momentum $p_{\rT,\Pb\Pbbar}$ of the
bottom--antibottom pair shown in \reffi{fi:distr-ptbb} looks very
similar.
Again, for our scale choice, the NLO corrections reduce the cross
section for large values of $p_{\rT,\Pb\Pbbar}$.

\begin{figure}
\includegraphics[bb= 85 440 285 660, width=.5\textwidth]{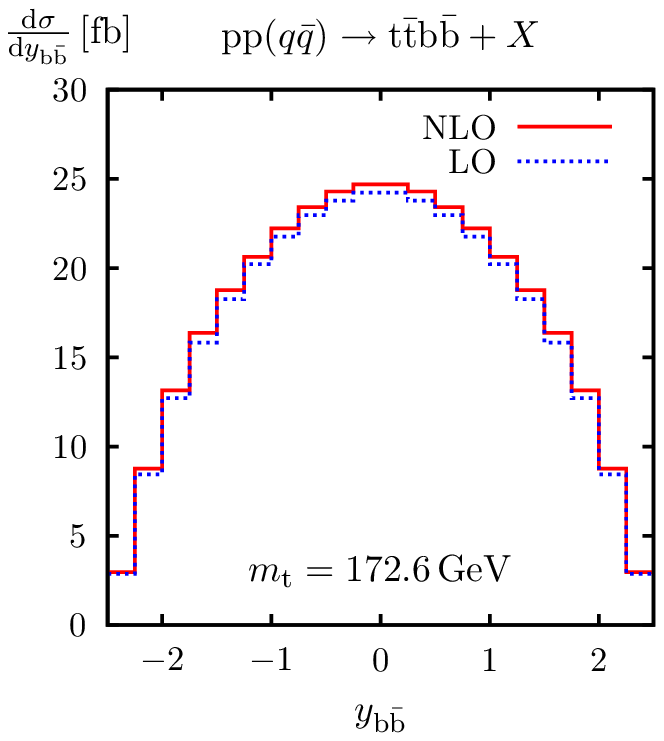}%
\includegraphics[bb= 85 440 285 660, width=.5\textwidth]{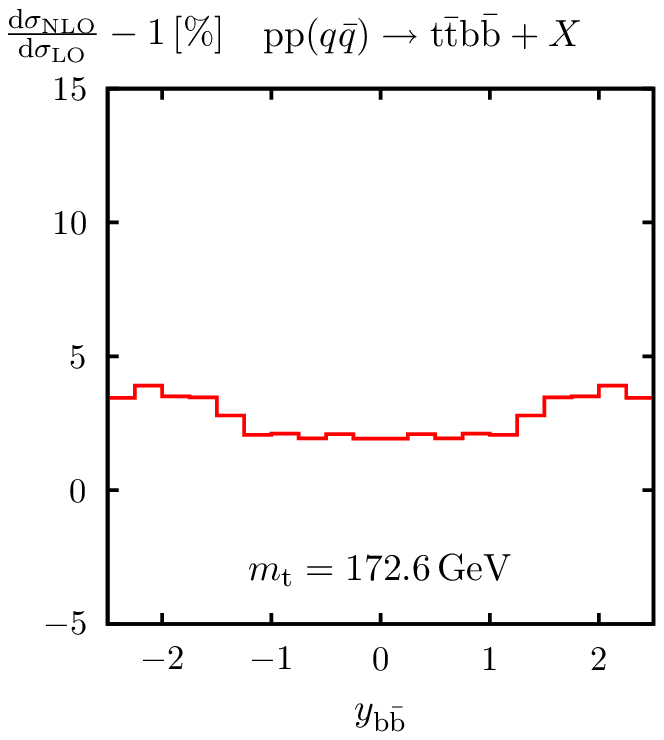}
\caption{Distribution in the rapidity $y_{\Pb\Pbbar}$ of the
  bottom--antibottom pair (left) and corresponding relative corrections
  (right) for $\mu_{\mathrm{R}}=\mu_{\mathrm{F}}=\mu_0=\Mt$.} 
\label{fi:distr-ybb}
\end{figure}

Finally, we depict the distribution in the rapidity $y_{\Pb\Pbbar}$ of
the bottom--antibottom pair in \reffi{fi:distr-ybb}.  
In this case,
the NLO corrections are rather flat at the level of 2.5\% with a
slight increase in the backward and forward directions.

\section{Conclusions}

Predictions for the background process
$\Pp\Pp\to\Pt\bar\Pt\Pb\bar\Pb+X$ in NLO QCD are indispensable for a
thorough analysis of $\Pt\bar\Pt\PH$ production at the LHC.

In this paper we have made the first step towards the full NLO
calculation upon evaluating the contribution from quark--antiquark
annihilation. We made use of the Feynman-diagrammatic approach
augmented by recently developed reduction techniques for one-loop
tensor integrals. We have devoted an appendix to the general issue of
rational terms resulting from ultraviolet or infrared (soft or
collinear) singularities within dimensional regularization.  In
particular, we have shown that rational terms of infrared origin
cancel in truncated one-loop diagrams for arbitrary processes in the
Feynman gauge and thus result only from wave-function renormalization.
Based on this observation we have formulated a general recipe for the
determination of rational terms in one-loop amplitudes.

Our calculation demonstrates that the Feynman-diagrammatic approach
can be successfully applied in the context of six-particle processes
at the LHC, providing excellent numerical stability and high speed.
The CPU time needed to evaluate the full virtual corrections with a
single processor is of the order of $10^{-2}$ seconds per phase-space
point.  Based on these encouraging results, we expect to be able to
extend this calculation to the technically more challenging
gluon-fusion channel.

\section*{Acknowledgements}
This work is supported in part by the European Community's Marie-Curie
Research Training Network under contract MRTN-CT-2006-035505 ``Tools
and Precision Calculations for Physics Discoveries at Colliders''.
A.~B.~would like to acknowledge support from the Japan Society for the
Promotion of Science (JSPS).

\appendix
\section*{Appendix}

\section{Rational terms in one-loop amplitudes}
\label{app:rat-terms}

In this appendix we elaborate on the issue of so-called rational terms
that occur within dimensional regularization when algebraic factors
depending on the space--time dimensionality $D$ multiply loop
integrals that contain UV or IR (soft and/or collinear) singularities,
which give rise to poles in $D-4$.
The calculation of rational terms is of particular importance for
so-called unitarity or generalized-unitarity methods (see
\citeres{Bern:2007dw,Ellis:2007br,Britto:2008vq,Berger:2008sj} and
references therein). In this context, these terms are typically
obtained by recursion relations
\cite{Bern:2005cq,Berger:2006ci,Badger:2007si} or by exploiting the
full $D$-dimensional dependence of the tree amplitudes
\cite{Anastasiou:2006gt,Britto:2008vq,Giele:2008ve}. Explicit recipes
to derive rational terms in the context of specific methods, which
make use of loop integrals in shifted space--time dimensions or employ
a numerical reduction at the integrand level, can be found in
\citeres{Binoth:2006hk,Ossola:2008xq}.  Here we discuss rational terms
in the framework of one-loop calculations that employ an arbitrary set
of tensor (and scalar) loop integrals in $D=4-2\eps$ dimensions and
derive general properties that are independent of the explicit
algorithm used for tensor reduction.

Specifically we investigate rational terms resulting from
unrenormalized truncated loop amplitudes and do not consider external
self-energy corrections (wave-function renormalization constants).
Since the latter enter via derivatives, our arguments cannot be
applied, but these contributions are easily calculated once and for
all.  We classify the different situations in which rational terms
arise and describe simple procedures and results for their actual
calculation.  In particular, we demonstrate that---for any scattering
amplitude involving quarks and gluons---the rational terms originating
from IR poles cancel within individual Feynman diagrams.  This
important property implies that, after separating the rational terms
of UV type, the coefficients of all IR-divergent tensor $N$-point
integrals can be evaluated in four dimensions.  In practice the
wave-function renormalization constants represent the only source of
rational terms of IR origin.  This greatly simplifies the algebraic
manipulation of IR-divergent scattering amplitudes.

\subsection{Classification of rational terms}

Algebraic factors containing the dimensionality $D$ 
result from two different 
sources in one-loop amplitudes:
\begin{enumerate}
\item ``Trace-like'' contractions among metric tensors or with Dirac
  structures lead to expressions such as $g^{\mu\nu}g_{\nu\mu}=D$,
  $\ga^\mu\dsl{a}\ga^\nu g_{\mu\nu}= \ga^\mu\dsl{a}\ga_\mu =
  (2-D)\dsl{a}$, etc.
\item In the reduction of tensor one-loop integrals to standard scalar
  integrals (such as the usual Passarino--Veltman
  reduction~\cite{Passarino:1979jh}) the tensor coefficients are
  eventually obtained as linear combinations of the scalar integrals
  which form a basis of functions.  In such linear combinations, the
  tensor coefficients containing metric tensors in their corresponding
  covariants receive prefactors with a dependence on $D$.

\end{enumerate}
Thus, we can distinguish four different types of rational terms,
classified according to type~1 or 2 being of UV or IR origin.

We employ the notation of \citere{Denner:2005nn}, where the covariant
coefficients of $N$-point integrals with rank $R$ are denoted as
$T^{N}_{i_1\dots i_R}$.  It is convenient to treat the UV- and
IR-divergent parts of tensor integrals separately. To this end, we
write
\beq\label{UVsing}
 T^{N}_{i_1\dots i_R}= \hat T^{N}_{i_1\dots i_R}
+ \frac{R^N_{i_1\dots i_R}}{\eps_{\mathrm{UV}}},
\eeq
where $R^N_{i_1\dots i_R}$ represents the (IR-finite) residue of the
UV pole of $T^{N}_{i_1\dots i_R}$, and $\hat T^{N}_{i_1\dots i_R}$ is
free from UV divergences but can contain single and double poles in
$\eps_{\mathrm{IR}}$ resulting from soft or collinear divergences.
The only IR-divergent 2-point functions are those without a scale.
These represent a special case since they vanish as a result of
cancellations between IR and UV poles, \ie they are formally
proportional to $(1/\eps_{\mathrm{UV}}-1/\eps_{\mathrm{IR}})$.  In
order to separate these UV and IR poles, we isolate the UV divergences
by writing, in the notation of \citere{Denner:2005nn},
\beq\label{UVsingB000}
B_{\underbrace{\scriptstyle0\dots0}_{m}
\underbrace{\scriptstyle1\dots1}_{n}}(0,0,0) = 
\hat B_{\underbrace{\scriptstyle0\dots0}_{m}
\underbrace{\scriptstyle1\dots1}_{n}}(0,0,0) +
\frac{(-1)^n}{n+1} 
\frac{\delta_{m0}}{\eps_{\mathrm{UV}}}.
\label{eq:B000}
\eeq
The UV-subtracted part is IR divergent,
\beq\label{IRsingB000}
\hat B_{\underbrace{\scriptstyle0\dots0}_{m}
\underbrace{\scriptstyle1\dots1}_{n}}(0,0,0) = 
-\frac{(-1)^n}{n+1} 
\frac{\delta_{m0}}{\eps_{\mathrm{IR}}}
,
\eeq
and exactly cancels against the UV pole.  However, we do no set
scaleless 2-point integrals to zero and treat the rational terms
resulting from $1/\eps_{\mathrm{UV}}$ and $\hat B_{\dots}(0,0,0)$
separately.
Since scaleless 2-point functions require light-like momentum
transfer ($p^2=0$), such integrals only occur in external
self-energy corrections, i.e.\ in wave-function renormalization constants, 
and in  the reduction of higher-point functions ($N>2$).
In the latter case, as we will show below,
the rational terms of IR origin cancel out.

\subsection{Rational terms of UV origin}
\label{app:UVrat}

\begin{sloppypar}
  The residues $R^N_{i_1\dots i_R}$ of the UV poles of general
  one-loop tensor integrals are simple polynomials of the external
  momenta and their explicit form is well known (see, e.g., App.~C of
  \citere{Denner:2002ii} and App.~A of \citere{Denner:2005nn}).  In
  particular, in contrast to IR divergences, the UV poles do not
  depend on kinematical properties of the amplitude such as on-shell
  relations of momenta.  This renders rational terms of UV origin very
  simple: Terms of type~2 can be directly included in the tensor
  reduction in a generic way, as e.g.\ done in
  \citeres{Denner:2002ii,Denner:2005nn,Passarino:1979jh,Denner:1991kt},
  and terms of type~1 can be extracted during the algebraic reduction
  of each Feynman diagram by means of a trivial expansion,
\beq\label{UVexpansion}
f(D) T^{N}_{i_1\dots i_R}= f(D)\hat T^{N}_{i_1\dots i_R} + \left[\frac{f(4)}{\eps_{\mathrm{UV}}}-2f'(4)\right] {R^N_{i_1\dots i_R}}.
\eeq
In the following we discuss the remaining rational terms that result 
from $f(D)\hat T^{N}_{i_1\dots i_R}$ when  $\hat T^{N}_{i_1\dots i_R}$
contains poles of IR origin.
\end{sloppypar}

\subsection{Rational terms of IR origin}

IR divergences of one-loop integrals are more complicated than UV
singularities, since they depend on specific kinematical properties of
amplitudes. According to Kinoshita~\cite{Kinoshita:ur} they can be
classified into soft and collinear singularities as indicated in
\reffi{fig:IRtops}:
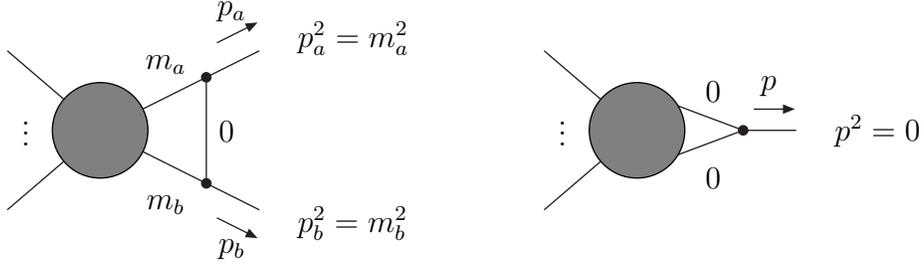
\begin{figure}
\centerline{
\begin{picture}(160,100)
\Line(40,50)(100,20)
\Line(40,50)(100,80)
\Line(80,70)(80,30)
\Vertex(80,70){2}
\Vertex(80,30){2}
\Line(40,50)(5,20)
\Line(40,50)(5,80)
\GCirc(40,50){18}{.5}
\Text(12,52)[c]{$\vdots$}
\Text(65,75)[c]{$m_a$}
\Text(88,50)[c]{$0$}
\Text(65,22)[c]{$m_b$}
\LongArrow(84,17)(98,10)
\LongArrow(84,83)(98,90)
\Text(90,95)[c]{$p_a$}
\Text(90, 5)[c]{$p_b$}
\Text(115,85)[l]{$p_a^2=m_a^2$}
\Text(115,15)[l]{$p_b^2=m_b^2$}
\end{picture}
\hspace*{3em}
\begin{picture}(160,100)
\Line(40,35)(80,50)
\Line(40,65)(80,50)
\Line(80,50)(100,50)
\Vertex(80,50){2}
\Line(40,50)(5,20)
\Line(40,50)(5,80)
\GCirc(40,50){18}{.5}
\Text(12,52)[c]{$\vdots$}
\Text(69,65)[c]{$0$}
\Text(69,32)[c]{$0$}
\LongArrow(84,58)(98,58)
\Text(90,67)[c]{$p$}
\Text(115,50)[l]{$p^2=0$}
\end{picture}
}
\caption{Kinematical configurations for soft (left) and collinear
(right) IR singularities in one-loop diagrams.}
\label{fig:IRtops}
\end{figure}
\begin{itemize}
\item A {\it soft} singularity arises if a massless particle is
  exchanged between two on-shell particles (see l.h.s.\ of
  \reffi{fig:IRtops}).  The singularity is logarithmic and originates
  from the region in momentum space where the momentum transfer of the
  massless propagator tends to zero.
\item A {\it collinear} singularity arises if an external line with a
  light-like momentum (e.g.\ a massless external on-shell particle) is
  attached to two massless propagators (see r.h.s.\ of
  \reffi{fig:IRtops}).  The singularity is also logarithmic and
  originates from the region in momentum space where the loop momentum
  of the two massless propagators becomes collinear to the momentum
  $p$ of the external particle.
\end{itemize}

We first consider rational terms of IR origin that can result from the
reduction of tensor integrals (type~2).  As can, for instance, be seen
from the results of \citere{Denner:2005nn}, only tensor coefficients
$T^{N}_{00\dots}$ whose covariants involve the metric tensor can get
$D$-dependent prefactors in reduction identities and could, thus, lead
to rational terms of type~2.  In Section~5.8 of \citere{Denner:2005nn}
it was, however, shown that these tensor coefficients are IR finite.
Thus, no type~2 rational terms of IR origin can result at all.

Moreover, reparametrizations of tensor integrals resulting from shifts
of the loop momentum and permutations of the propagators do not give
rise to $D$-dependent factors or relations between (IR-finite) tensor
coefficients of type $T^{N}_{00\dots}$ and IR-singular integrals.
Therefore, in order to demonstrate that a certain diagram is free from
IR rational terms it is sufficient to find for each soft- or
collinear-singular region a specific representation that is manifestly
free from IR rational terms of type~1, \ie to find an expression in
which the corresponding IR-divergent part is expressed as a linear
combination of IR-divergent tensor (or scalar) integrals with
coefficients that are independent of $D$.  In explicit calculations,
this can be achieved by means of an algebraic reduction that
implements all possible relations between the IR-divergent parts of
tensor integrals.  This task is non-trivial since the standard scalar
integrals do not provide a unique representation of IR divergences.
For instance, IR-finite parts of a diagram can be expressed as linear
combinations of IR-divergent 4-point and 3-point scalar integrals.
More generally, IR-singular $N$-point tensor integrals can be
expressed in terms of IR-divergent 3- and 2-point scalar integrals
plus IR-finite terms. This reduction of IR singularities is explicitly
implemented in the algorithm presented in \citere{Dittmaier:2003bc}
and can be summarized by the following formula (see Eq.~(3.14) in
\citere{Dittmaier:2003bc}) which relates the IR-divergent part of
$N$-point tensor integrals to 3-point tensor integrals associated with
the IR-divergent triangle subdiagrams,
\beqar\label{IRseparation}
\lefteqn{
T^{N}_{\mu_1\dots \mu_R}(p_0,\dots,p_{N-1},m_0,\dots,m_{N-1})
}\quad&&\nl
&=&
\sum_{n=0}^{N-1}
\sum_{k=0\atop k\neq n,n+1}^{N-1}
A_{nk}C_{\mu_1\dots \mu_R}(p_n,p_{n+1},p_{k},m_n,m_{n+1},m_k)
+\mbox{IR-finite part}.
\eeqar
The coefficients $A_{nk}$ are independent of $D$ since all relations
between IR-divergent tensor integrals are free from $D$-dimensional
coefficients. For the explicit form of $A_{nk}$ and details of the
notation we refer to \citere{Dittmaier:2003bc}.  In practice, using
\refeq{IRseparation} and performing a subsequent reduction to scalar
integrals, one can construct a unique representation of IR divergences
in terms of 3- and 2-point scalar functions. The 3-point functions on
the right-hand side of \refeq{IRseparation} can be subtracted from the
complete tensor integral leading to an IR-finite expression which can
be evaluated in 4 dimensions. Thus, only the re-added 3-point
functions \refeq{IRseparation} have to be manipulated in $D$
dimensions.

Using this approach, we have observed in explicit calculations for
$\Pp\Pp\to\Pt\bar\Pt\PH$ \cite{Beenakker:2002nc},
$\Pp\Pp\to\Pt\bar\Pt+\mathrm{jet}$ \cite{Dittmaier:2007wz}, and
$\Pp\Pp\to\PW\PW+\mathrm{jet}$ \cite{Dittmaier:2007th} that---after
complete reduction of the IR divergences---the coefficients of the
IR-singular $C_0$ and $B_0$ functions are independent of $D$, \ie that
rational terms of IR type cancel completely.  This indicates, {\it a
  posteriori}, that the terms $f(D)\hat T^{N}_{i_1\dots i_R}$ in
\refeq{UVexpansion} can be replaced by $f(4)\hat T^{N}_{i_1\dots i_R}$
{\it from the beginning} in the calculation.
In the following we demonstrate that, in the Feynman gauge, the
cancellation of IR rational terms is a general property of scattering
amplitudes involving an arbitrary number of external quarks and
gluons.

To this end, we inspect the integrand of a general one-loop
IR-divergent diagram in momentum space and, in the spirit of
\citere{Dittmaier:2003bc}, we separate the IR singularities associated
with different soft and collinear regions and relate them to triangle
subdiagrams.  Using on-shell relations we show that, in the soft and
collinear regions, the integrands can be cast into a form that is free
from ``trace-like" contractions, which potentially produce
$D$-dependent factors.  In this way we obtain a generic representation
of the IR singularities that is manifestly free from rational terms of
IR type.

The considerations presented in the following are not new.  A similar
approach is used, for instance, in \citere{Nagy:2003qn}, where IR and
UV singularities are subtracted at the diagrammatic level and isolated
in simple process-independent terms in order to obtain numerically
integrable expressions.  Using similar techniques, here we consider
soft and collinear contributions from the perspective of an analytic
calculation in terms of divergent one-loop tensor and scalar integrals
and discuss the rational terms associated with IR divergences.

\paragraph{Soft singularities:}

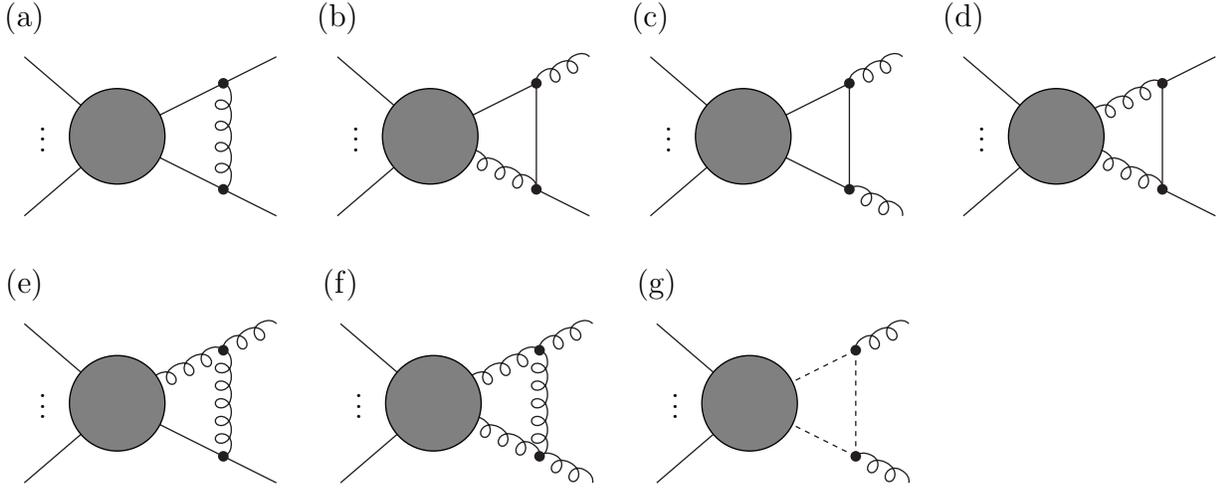
\begin{figure}
\begin{picture}(100,100)
\Line(40,50)(100,20)
\Line(40,50)(100,80)
\Gluon(80,70)(80,30){3}{4}
\Vertex(80,70){2}
\Vertex(80,30){2}
\Line(40,50)(5,20)
\Line(40,50)(5,80)
\GCirc(40,50){18}{.5}
\Text(12,52)[c]{$\vdots$}
\Text(5,95)[c]{(a)}
\end{picture}
\hspace*{1em}
\begin{picture}(100,100)
\Line(40,50)( 80,70)
\Gluon(80,70)(100,80){3}{2.5}
\Gluon(40,50)( 80,30){3}{5}
\Line(80,30)(100,20)
\Line(80,70)(80,30)
\Vertex(80,70){2}
\Vertex(80,30){2}
\Line(40,50)(5,20)
\Line(40,50)(5,80)
\GCirc(40,50){18}{.5}
\Text(12,52)[c]{$\vdots$}
\Text(5,95)[c]{(b)}
\end{picture}
\hspace*{1em}
\begin{picture}(100,100)
\Line(40,50)( 80,70)
\Gluon(80,70)(100,80){3}{2.5}
\Line(40,50)( 80,30)
\Gluon(80,30)(100,20){3}{2.5}
\Line(80,70)(80,30)
\Vertex(80,70){2}
\Vertex(80,30){2}
\Line(40,50)(5,20)
\Line(40,50)(5,80)
\GCirc(40,50){18}{.5}
\Text(12,52)[c]{$\vdots$}
\Text(5,95)[c]{(c)}
\end{picture}
\hspace*{1em}
\begin{picture}(100,100)
\Gluon(40,50)( 80,70){3}{5}
\Line(80,70)(100,80)
\Gluon(40,50)( 80,30){3}{5}
\Line(80,30)(100,20)
\Line(80,70)(80,30)
\Vertex(80,70){2}
\Vertex(80,30){2}
\Line(40,50)(5,20)
\Line(40,50)(5,80)
\GCirc(40,50){18}{.5}
\Text(12,52)[c]{$\vdots$}
\Text(5,95)[c]{(d)}
\end{picture}
\\
\begin{picture}(100,100)
\Gluon(40,50)( 80,70){3}{5}
\Gluon(80,70)(100,80){3}{2.5}
\Line(40,50)( 80,30)
\Line(80,30)(100,20)
\Gluon(80,70)(80,30){3}{5}
\Vertex(80,70){2}
\Vertex(80,30){2}
\Line(40,50)(5,20)
\Line(40,50)(5,80)
\GCirc(40,50){18}{.5}
\Text(12,52)[c]{$\vdots$}
\Text(5,95)[c]{(e)}
\end{picture}
\hspace*{1em}
\begin{picture}(100,100)
\Gluon(40,50)( 80,70){3}{5}
\Gluon(80,70)(100,80){3}{2.5}
\Gluon(40,50)( 80,30){3}{5}
\Gluon(80,30)(100,20){3}{2.5}
\Gluon(80,70)(80,30){3}{5}
\Vertex(80,70){2}
\Vertex(80,30){2}
\Line(40,50)(5,20)
\Line(40,50)(5,80)
\GCirc(40,50){18}{.5}
\Text(12,52)[c]{$\vdots$}
\Text(5,95)[c]{(f)}
\end{picture}
\hspace*{1em}
\begin{picture}(100,100)
\DashLine(40,50)( 80,70){2}
\Gluon(80,70)(100,80){3}{2.5}
\DashLine(40,50)( 80,30){2}
\Gluon(80,30)(100,20){3}{2.5}
\DashLine(80,70)(80,30){2}
\Vertex(80,70){2}
\Vertex(80,30){2}
\Line(40,50)(5,20)
\Line(40,50)(5,80)
\GCirc(40,50){18}{.5}
\Text(12,52)[c]{$\vdots$}
\Text(5,95)[c]{(g)}
\end{picture}
\vspace*{-1em}
\caption{Subdiagrams containing soft-singular integrals, with solid lines
indicating quarks, epicycles gluons, and dotted lines ghosts.}
\label{fig:softdiags}
\end{figure}

Figure~\ref{fig:softdiags} shows all potentially soft-divergent
subdiagrams containing quarks, gluons, or Faddeev--Popov ghosts.
Since the soft singularity is related to zero-momentum transfer on the
internal line linking the two external particles, it is convenient to
identify the integration momentum $q$ of the loop integral with the
momentum on this line. Since the soft singularity is logarithmic, all
contributions of $q$ in the numerator of the integral are IR finite.
In other words, being only after the IR-divergent part we can set $q$
to zero in the numerator and in propagators that do not belong to the
soft-singular triangle subdiagram. This procedure immediately kills
the three subintegrals (b)--(d) with a quark on the $q$ line because
of the factor $\dsl{q}$ in the (massless) quark propagator.  Diagram
(g) with a ghost coupling to external gluons does not contribute
either, because the integrand receives factors (depending on the
gauge) of $q\veps_a\to0$ or $p_a\veps_a=0$ from the coupling of the
ghost to the on-shell gluon with momentum $p_a$ and polarization
vector $\veps_a$.  The amplitude $\M_{(8\mathrm{a})}$ of diagram (a),
of course, involves a soft singularity, but without any $D$
dependence, as can be seen in the following example where we consider
an outgoing quark--antiquark pair,
\beqar
\M_{(8\mathrm{a})} &=& 
\int \rd^D q\; \frac{g^{\mu\nu}}{q^2} \,
\bar u_a(p_a) \, \gamma_\mu \,
\frac{\dsl{q}+\dsl{p}_a+m_a}{(q+p_a)^2-m_a^2} \, \Ga(q) \,
\frac{\dsl{q}-\dsl{p}_b+m_b}{(q-p_b)^2-m_b^2} \,
\gamma_\nu \, v_b(p_b)
\nn\\
&=&
\int \rd^D q\; \frac{g^{\mu\nu}}{q^2} \,
\bar u_a(p_a) \, \gamma_\mu \,
\frac{\dsl{p}_a+m_a}{(q+p_a)^2-m_a^2} \, \Ga(0) \,
\frac{-\dsl{p}_b+m_b}{(q-p_b)^2-m_b^2} \,
\gamma_\nu \, v_b(p_b)
+\dots
\nn\\
&=&
-4(p_a p_b)
\int \rd^D q\; \frac{\bar u_a(p_a) \, \Ga(0) \, v_b(p_b)
}{q^2[(q+p_a)^2-m_a^2][(q-p_b)^2-m_b^2]} 
+\dots\,.
\label{eq:M8a}
\eeqar
The Dirac structure $\Ga(q)$ contains the remaining part of the
diagram and the ellipses stand for terms that are not singular if the
gluon becomes soft.  For the last equality the Dirac equation was used
twice. The soft singularity, which is just contained in the scalar
3-point function, does not receive $D$-dependent factors and, thus,
does not deliver rational terms, because $\Ga(0)$ is a tree-like
structure and does not contain a trace-like contraction that would
lead to factors of $D$. Such factors, e.g., arise in terms like
$\ga^\mu\Ga(0)\ga_\mu$, which are absent in the soft-singular part.
The same reasoning applies also to all other possible fermion-number
flows in diagram (a).  The remaining two diagrams (e) and (f) of
\reffi{fig:IRtops} can be analysed in the same way, leading to the
same conclusion that no $D$-dependent factors multiply soft-singular
integrals.  For brevity we show this only for diagram (f) explicitly,
\beqar
\M_{(8\mathrm{f})} &=& 
\int \rd^D q\; \frac{g^{\la\si}}{q^2} \,
\veps^{\mu*}_a \,
\frac{g_{\mu\nu}(q+2p_a)_\la-g_{\nu\la}(2q+p_a)_\mu+g_{\la\mu}(q-p_a)_\nu}
{(q+p_a)^2} \, \Ga^{\nu\tau}(q) 
\nn\\
&& {} \qquad \times
\veps^{\rho*}_b \,
\frac{g_{\rho\tau}(-q+2p_b)_\si+g_{\tau\si}(2q-p_b)_\rho
-g_{\si\rho}(q+p_b)_\tau}
{(q-p_b)^2}
\nn\\
&=&
\int \rd^D q\; \frac{g^{\la\si}}{q^2} \,
\veps^{\mu*}_a
\frac{2g_{\mu\nu}p_{a,\la}-g_{\nu\la}p_{a,\mu}-g_{\la\mu}p_{a,\nu}}
{(q+p_a)^2} \, \Ga^{\nu\tau}(0) 
\nn\\
&& {} \qquad \times
\veps^{\rho*}_b
\frac{2g_{\rho\tau}p_{b,\si}-g_{\tau\si}p_{b,\rho}-g_{\si\rho}p_{b,\tau}}
{(q-p_b)^2} + \dots
\nn\\
&=&
\int \rd^D q\; 
\frac{
[4(p_ap_b) \veps^*_{a,\nu}\veps^*_{b,\tau}
-2(\veps^*_a p_b) p_{a,\nu}\veps^*_{b,\tau}
-2(p_a\veps^*_b) \veps^*_{a,\nu} p_{b,\tau}
+ (\veps^*_a \veps^*_b) p_{a,\nu} p_{b,\tau}]
\Ga^{\nu\tau}(0) }
{q^2(q+p_a)^2(q-p_b)^2} 
\nn\\
&& \qquad {} + \;\dots\,,
\label{eq:M8f}
\eeqar
where we assume the external on-shell gluons to be outgoing.  The last
line of this result cannot contain explicit factors of $D$, since
those would require a trace-like contraction ${\Ga^\nu}_\nu(0)$;
instead $\Ga$ is only contracted with momenta $p_{a,b}$ and
polarization vectors $\veps^*_{a,b}$. Formally the contraction
${\Ga^\nu}_\nu(0)$ occurs, but with a proportionality to
$(p_a\veps^*_a)(p_b\veps^*_b)$ which vanishes owing to the on-shell
condition of the gluons.

It is well known that the soft singularities are ruled by the eikonal
current, with the result that divergences connected to soft-particle
exchange between $a$ and $b$ are proportional to $(p_a p_b)$. For
$\M_{(8\mathrm{a})}$ this factor is already explicit in
\refeq{eq:M8a}, for $\M_{(8\mathrm{f})}$ in \refeq{eq:M8f} this fact
becomes obvious after making use of the Ward identities
$p_{a,\nu}\veps^*_{b,\tau} \Ga^{\nu\tau}(0) =\veps^*_{a,\nu}
p_{b,\tau} \Ga^{\nu\tau}(0) =p_{a,\nu} p_{b,\tau} \Ga^{\nu\tau}(0)=0$,
which are valid if all other external particles are on shell and all
diagrams contributing to $\Ga$ are summed over.

\paragraph{Collinear singularities:}

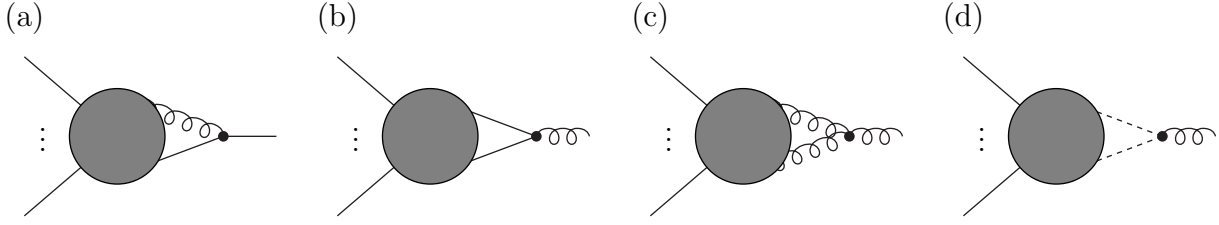
\begin{figure}
\begin{picture}(100,100)
\Line(40,35)(80,50)
\Gluon(40,65)(80,50){3}{5}
\Line(80,50)(100,50)
\Vertex(80,50){2}
\Line(40,50)(5,20)
\Line(40,50)(5,80)
\GCirc(40,50){18}{.5}
\Text(12,52)[c]{$\vdots$}
\Text(5,95)[c]{(a)}
\end{picture}
\hspace*{1em}
\begin{picture}(100,100)
\Line(40,35)(80,50)
\Line(40,65)(80,50)
\Gluon(80,50)(100,50){3}{2.5}
\Vertex(80,50){2}
\Line(40,50)(5,20)
\Line(40,50)(5,80)
\GCirc(40,50){18}{.5}
\Text(12,52)[c]{$\vdots$}
\Text(5,95)[c]{(b)}
\end{picture}
\hspace*{1em}
\begin{picture}(100,100)
\Gluon(40,35)(80,50){3}{5}
\Gluon(40,65)(80,50){3}{5}
\Gluon(80,50)(100,50){3}{2.5}
\Vertex(80,50){2}
\Line(40,50)(5,20)
\Line(40,50)(5,80)
\GCirc(40,50){18}{.5}
\Text(12,52)[c]{$\vdots$}
\Text(5,95)[c]{(c)}
\end{picture}
\hspace*{1em}
\begin{picture}(100,100)
\DashLine(40,35)(80,50){2}
\DashLine(40,65)(80,50){2}
\Gluon(80,50)(100,50){3}{2.5}
\Vertex(80,50){2}
\Line(40,50)(5,20)
\Line(40,50)(5,80)
\GCirc(40,50){18}{.5}
\Text(12,52)[c]{$\vdots$}
\Text(5,95)[c]{(d)}
\end{picture}
\caption{Subdiagrams containing collinear-singular integrals, with solid lines
indicating quarks, epicycles gluons, and dotted lines ghosts.}
\label{fig:colldiags}
\end{figure}

Figure~\ref{fig:colldiags} shows all potentially collinear-divergent
subdiagrams containing (massless) 
quarks, gluons, or Faddeev--Popov ghosts.
We identify the integration momentum $q$ with the momentum of one of
the two propagators attached to the external light-like line, which
carries the momentum $p$ ($p^2=0$). The collinear singularity stems
from the region in $q$ space where $q$ is collinear to $p$, i.e.\ 
$q^\mu=x(q)p^\mu$ with some scalar function $x(q)$. Since the
singularity is only logarithmic, we do not change the singularity if
we replace $q^\mu$ by $x(q)p^\mu$ in the numerator of the amplitude.
The on-shell conditions (Dirac equation for quarks and transversality
condition for gluons) then imply that the collinear-singular parts of
diagrams (a)--(d) in \reffi{fig:colldiags} do not receive
$D$-dependent factors:
\beqar
\M_{(9\mathrm{a})} &=& 
\int \rd^D q\; \frac{1}{q^2} \,
\bar u(p) \, \gamma_\mu \,
\frac{\dsl{q}+\dsl{p}}{(q+p)^2} \, \Ga^\mu(q) 
\nn\\
&=&
\int \rd^D q\; \frac{1}{q^2} \,
\bar u(p) \, \gamma_\mu \,
\frac{[1+x(q)]\dsl{p}}{(q+p)^2} \, \Ga^\mu(x(q)p) + \dots
\nn\\
&=&
\int \rd^D q\; 
\frac{ 2[1+x(q)] \bar u(p) p_\mu \Ga^\mu(x(q)p) 
}{q^2(q+p)^2} \,
+ \dots\,,
\\
\M_{(9\mathrm{b})} 
&=& 
\int \rd^D q\; \frac{1}{q^2 (q+p)^2}\, 
{\left[\dsl{q} \, \dsl{\veps}^* \, 
(\dsl{q}+\dsl{p})\right]_{\alpha\beta}}
\, \Ga_{\alpha\beta}(q)
\nn\\
&=& 
\int \rd^D q\; \frac{x(q)[1+x(q)]}{q^2 (q+p)^2}\, 
{\left[\dsl{p} \, \dsl{\veps}^* \, 
\dsl{p}\right]_{\alpha\beta}}
\, \Ga_{\alpha\beta}(x(q)p) + \dots
\nn\\
&=& 
\int \rd^D q\; \frac{x(q)[1+x(q)]}{q^2 (q+p)^2}\, 
{2(\veps^*p)\,\left[\dsl{p}\right]_{\alpha\beta}}
\, \Ga_{\alpha\beta}(x(q)p) + \dots
\nn\\
&=&0+\dots\,,
\\
\M_{(9\mathrm{c})} &=& 
\int \rd^D q\; 
\veps^{\mu*} \,
\frac{g_{\mu\nu}(q+2p)_\la-g_{\nu\la}(2q+p)_\mu+g_{\la\mu}(q-p)_\nu}
{q^2(q+p)^2} \, \Ga^{\nu\la}(q) 
\nn\\
&=&
\int \rd^D q\; 
\veps^{\mu*} \,
\frac{g_{\mu\nu}(x(q)+2)p_\la-g_{\nu\la}(2x(q)+1)p_\mu+g_{\la\mu}(x(q)-1)p_\nu}
{q^2(q+p)^2} \, \Ga^{\nu\la}(x(q)p) + \dots
\nn\\
&=&
\int \rd^D q\; 
\frac{[(x(q)+2)\veps^*_\nu p_\la+(x(q)-1)\veps^*_\la p_\nu] \, 
\Ga^{\nu\la}(x(q)p)}
{q^2(q+p)^2} + \dots\,,
\\
\M_{(9\mathrm{d})} &=& 
\int \rd^D q\; 
\veps^{\mu*} \, \frac{q_\mu} {q^2(q+p)^2} \, \Ga(q) 
\nn\\
&=&
\int \rd^D q\; 
\veps^{\mu*} \, \frac{x(q)p_\mu} {q^2(q+p)^2} \, \Ga(x(q)p) 
\nn\\
&=&
0+\dots\,.
\eeqar
Note that diagrams (b) and (d) do not have collinear singularities at
all.  The collinear divergences in diagrams (a) and (c) do not receive
$D$-dependent factors since, in the collinear region, the $\Ga$ terms
do not take part in trace-like contractions like $\ga_\mu\Ga^\mu$ or
${\Ga^\nu}_\nu$, and terms $q^{\tau_1}\dots q^{\tau_n}$ inside $\Ga$
yield tensor structures $[x(q)]^n p^{\tau_1}\dots p^{\tau_n}$ without
metric tensors.  The above IR-divergent integrals are easily expressed
in terms of usual 3-point functions by observing that in the collinear
limit all propagators inside $\Ga(x(q)p)$ are linear in $x(q)$ owing
to $(q+p_k)^2\to 2 x(q) p p_k+p_k^2$.  Thus, one can easily express
the denominator of $\Ga(x(q)p)$ as a linear combination of single
propagators via partial fractioning, as done in
\citere{Dittmaier:2003bc}.

In summary we have obtained a representation of the soft and collinear
singularities of generic Feynman diagrams in terms of 3-point tensor
integrals that are explicitly free from $D$-dimensional prefactors.
We can, thus, conclude that rational terms of IR origin cancel in any
unrenormalized scattering amplitude and can be neglected {\it a
  priori} in explicit calculations.
This property is a consequence of the Lorentz structure of the gluon
couplings and the logarithmic nature of IR singularities within the
conventional Feynman gauge.  However, the cancellation of IR rational
terms holds also in more general gauge-fixings, as for instance the
background-field Feynman gauge (see \citere{Abbott:1980hw} and
references therein), where the Lorentz structures in the gluon
couplings and the poles of the propagators behave as in the Feynman
gauge.

We point out that the cancellation of IR rational terms is independent
of the actual reduction method employed in the calculation and is
generally valid in any approach where the IR-divergent parts of loop
diagrams are entirely expressed in terms of tensor (or scalar)
$N$-point integrals in $4-2\eps$ dimensions.

\subsection{A recipe for determining rational terms}

Based on the above considerations we can formulate a simple algorithm
for determining all rational terms of either UV or IR origin.  For
unrenormalized truncated one-loop diagrams, i.e.\ excluding
counterterm diagrams and self-energy corrections to external lines,
proceed as follows:
\begin{enumerate}
\item 
  Separate UV and IR divergences of all tensor integrals as in
  \refeq{UVsing}, thereby keeping track of the poles of all scaleless
  2-point integrals as indicated in \refeq{UVsingB000} and
  \refeq{IRsingB000}.
  
\item 
  Extract the rational terms of UV origin as described in
  \refeq{UVexpansion} including the rational terms resulting from UV
  poles of scaleless 2-point integrals.
  
\item 
  Ignore rational terms of IR origin upon replacing $f(D)\to f(4)$
  on the right-hand side of \refeq{UVexpansion}, because the arguments
  of the previous section show that all rational terms resulting from
  $[f(D)-f(4)]\hat T^{N}_{i_1\dots i_R}$ in \refeq{UVexpansion}
  compensate each other (even if they may arise in intermediate
  steps).
\end{enumerate}

This recipe does not apply to wave-function renormalization constants.
These can be easily calculated, and explicit results are,
e.g., given in Eqs.~(2.27)--(2.28) of \citere{Beenakker:2002nc}.

\section{Four-dimensional reduction of Dirac chains to standard matrix elements}
\label{app:SMEs}

\newcommand{\ps}{p\hspace{-0.42em}/}

Here we outline the algebraic procedure employed to reduce Dirac
structures to standard matrix elements.  The reduction is based on the
strategy described in Sect.~3.3 of \citere{Denner:2005fg}, which is
worked out for massless six-fermion processes, and involves additional
features to treat the Dirac chains associated with massive top quarks.
This method exploits a set of relations that do not give rise to
denominators involving kinematical variables, thereby avoiding
possible numerical instabilities in exceptional phase-space regions.
Refraining from a detailed description of the entire reduction
algorithm, which is quite involved, we only outline the basic
principles, which can be traced back to a few simple identities.

Tree and loop diagrams give rise to a large number of Dirac structures
of the type
\beqar\label{eq:diracstruct}
\bar v(p_1)
\gamma^{\mu_1} 
\ldots
{\ps_{i_1}}
\ldots
u(p_2)
\;
\bar v(p_3)
\gamma^{\nu_1} 
\ldots
{\ps_{j_1}}
\ldots
u(p_4)
\;
\bar v(p_5)
\gamma^{\rho_1} 
\ldots
{\ps_{k_1}}
\ldots
u(p_6)
,
\eeqar
which consist of gamma matrices (or slashed momenta) that are
sandwiched between the spinors $\bar{v}(p_a)$ and $u(p_{a+1})$ of the
six (anti)fermions.  For convenience we consider the crossed process
$\bar q q \bar\Pt \Pt \bar \Pb \Pb \to 0$, where all particles and
their momenta are incoming.  While the chains associated with massless
quarks ($a=1,5$) contain only odd numbers of Dirac matrices, inside
the top chain ($a=3$) also even numbers of Dirac matrices appear.  The
open Lorentz indices $\mu_i,\nu_j,\rho_k$ in \refeq{eq:diracstruct}
are always pairwise contracted via the metric tensor.

The combinations of Dirac matrices occurring inside individual chains
$\bar v(p_a)\dots u(p_{a+1})$ can easily be simplified by means of
elementary relations:
\begin{enumerate}
\renewcommand{\labelenumi}{(\roman{enumi})}
\item Standard Dirac algebra permits to reduce $\gamma^\mu\dots
  \gamma_\mu$ contractions inside Dirac chains, to bring gamma
  matrices and $\ps$ terms into a standard order through
  anti-com\-mu\-ta\-ti\-ons, and to eliminate $\ps^2$ terms.
\item 
  All $\ps_a$ and $\ps_{a+1}$ terms can be eliminated with
  the Dirac equation.
\item 
  The $\ps_i$ terms associated with one of the other
  external momenta ($i\neq a,a+1$) can be eliminated via momentum
  conservation.
\end{enumerate}

After these simplifications, which we perform in $D$ dimensions, we
are left with a still large number of Dirac structures of
$\mathcal{O}(10^3)$.
To obtain a further reduction we employ additional identities which
permit to shift $\ps$ terms and $\gamma^\mu$ matrices with open
indices from one chain to another, thereby permitting further
simplifications of type (i)--(iii).
This part of the reduction relies on four-dimensional relations and is
performed after separation of all $(D-4)$-poles in dimensional
regularization.

All four-dimensional identities are derived from basic relations which
follow from Chisholm's identity (see Eqs.~(3.4)--(3.6) in
\citere{Denner:2005fg}) and read
\beqar\label{eq:chisholm1}
{
\gamma^\mu}
{\gamma^\alpha
\gamma^\beta}
\omega_\pm
\otimes
{
\gamma_{\mu}}
&=&
{
\gamma^\mu}
\omega_\pm
\otimes
\left(
\gamma_{\mu}
\gamma^\beta
\gamma^\alpha
\omega_\pm
+
\gamma^\alpha
\gamma^\beta
\gamma_{\mu}
\omega_\mp
\right)
,\nl
{
\gamma^\alpha}
{\gamma^\mu
\gamma^\beta}
\omega_\pm
\otimes
{
\gamma_{\mu}}
&=&
{
\gamma^\mu}
\omega_\pm
\otimes
\left(
\gamma^{\beta}
\gamma_\mu
\gamma^\alpha
\omega_\pm
+
\gamma^\alpha
\gamma_\mu
\gamma^{\beta}
\omega_\mp
\right)
,\nl
{
\gamma^\alpha}
{\gamma^\beta
\gamma^\mu}
\omega_\pm
\otimes
{
\gamma_{\mu}}
&=&
{
\gamma^\mu}
\omega_\pm
\otimes
\left(
\gamma^{\beta}
\gamma^\alpha
\gamma_\mu
\omega_\pm
+
\gamma_\mu
\gamma^\alpha
\gamma^{\beta}
\omega_\mp
\right),
\eeqar
where $\omega_\pm=(1\pm\gamma^5)/2$ are the chirality projectors and
the tensor products connect different Dirac chains.  In order to
exploit these relations we introduce chirality projectors inside every
Dirac chain,
$
\bar{v}(p_a)
\Gamma
u(p_{a+1})
=
\sum_{\lambda=\pm}
\bar{v}(p_a)
\Gamma
\omega_\lambda
u(p_{a+1}).
$
Then we can use \refeq{eq:chisholm1} to exchange 
$\gamma^\alpha\gamma^\beta$-terms between chains that are
connected via  $\gamma^\mu$-contractions.
A simple application of \refeq{eq:chisholm1} is given by 
\beqar\label{eq:chisholm2}
{\gamma^\mu}
{\gamma^\alpha}
{\gamma^\nu}
\omega_\pm
\otimes
{\gamma_{\mu}}
{\gamma^\beta}
{\gamma_\nu}
=
4g^{\alpha\beta}
{\gamma^\mu}
\omega_\pm
\otimes
\gamma_\mu
\omega_\pm 
+
4
\gamma^\beta
\omega_\pm
\otimes
\gamma^\alpha
\omega_\mp 
,\nl
{\gamma^\mu}
{\gamma^\alpha}
{\gamma^\nu}
\omega_\pm
\otimes
{\gamma_{\nu}}
{\gamma^\beta}
{\gamma_\mu}
=
4g^{\alpha\beta}
{\gamma^\mu}
\omega_\pm
\otimes
\gamma_\mu
\omega_\mp 
+
4
\gamma^\beta
\omega_\pm
\otimes
\gamma^\alpha
\omega_\pm 
.
\eeqar
These identities permit to eliminate double Lorentz contractions
between two Dirac chains.  Alternatively we can use
\refeq{eq:chisholm2} in the opposite direction, in combination with
the Dirac equation.  This yields the relations
\beqar\label{eq:chisholm3}
\lefteqn{
\ps_b \omega_\pm u(p_a) \otimes \ps_a \omega_\mp u(p_b)
=
(p_a p_b)\gamma^\mu \omega_\pm u(p_a) \otimes \gamma_\mu 
\omega_\mp
u(p_b)}\quad&&
\nl&&{}
- \frac{m_a}{2}\gamma^\mu \ps_b \omega_\mp u(p_a) \otimes \gamma_\mu 
\omega_\mp u(p_b)
- \frac{m_b}{2}\gamma^\mu \omega_\pm u(p_a) \otimes \gamma_\mu 
\ps_a \omega_\pm u(p_b)
+m_am_b\mbox{-term},
\nn\\
\lefteqn{
\bar v(p_a)\ps_b \omega_\pm \otimes \bar v(p_b)\ps_a \omega_\mp 
=
(p_a p_b)\bar v(p_a)\gamma^\mu \omega_\pm \otimes \bar v(p_b)\gamma_\mu 
\omega_\mp}\quad&&
\nl&&{}
+ \frac{m_a}{2}\bar v(p_a)\ps_b\gamma^\mu  \omega_\pm \otimes \bar v(p_b) \gamma_\mu 
\omega_\mp 
+ \frac{m_b}{2}\bar v(p_a)\gamma^\mu \omega_\pm \otimes \bar v(p_b) \ps_a\gamma_\mu 
 \omega_\mp 
+m_am_b\mbox{-term},
\nn\\
\lefteqn{
\bar v(p_a)\ps_b \omega_\pm \otimes \ps_a \omega_\pm u(p_b)
=
(p_a p_b)\bar v(p_a)\gamma^\mu \omega_\pm \otimes \gamma_\mu 
\omega_\pm u(p_b)}\quad&&
\nl&&{}
+ \frac{m_a}{2}\bar v(p_a)\ps_b\gamma^\mu  \omega_\pm \otimes  \gamma_\mu 
\omega_\pm  u(p_b)
+ \frac{m_b}{2}\bar v(p_a)\gamma^\mu \omega_\pm \otimes  \ps_a \gamma_\mu
 \omega_\mp  u(p_b),
\eeqar
where the terms proportional to $m_am_b$ vanish since the
(anti)spinors associated with particles $a$ and $b$ belong to
different Dirac chains and, thus, at least one of them is massless in
our case.  The relations \refeq{eq:chisholm3} can be used to reduce
the number of $\ps_i$ terms in the Dirac chains.

We give two explicit examples to illustrate the four-dimensional
reduction of Dirac structures that involve one massive Dirac chain:
\begin{itemize}
\item[] {\it Example 1:} \\
  We reduce terms involving double contractions of the type
  $\gamma^\mu \gamma^\nu \otimes \gamma_\mu \otimes \gamma_\nu$ to
  structures involving only single contractions of Lorentz indices as
  much as possible.  This procedure somewhat generalizes Step~1 in
  Section~3.3 of \citere{Denner:2005fg}.  Following the notation of
  that reference, we use the shorthand $\left[\Gamma\right]^\rho_{ij}=
  \bar v(p_i) \Gamma \omega_\rho u(p_j)$ and consider Dirac structures
  of the type
\beqar\label{doublecontractions}
\left[A_0\gamma^\mu\gamma^\nu\right]^\rho_{ij}
\left[A_1\gamma_\mu\right]^\sigma_{kl}
\left[A_2\gamma_\nu\right]^\tau_{mn},
\eeqar
where the terms $A_{i}$ consist of $\ps$-products, each containing
$n_{i}$ slashed momenta.  By means of the following two steps the
structures \refeq{doublecontractions} can be recursively reduced to
$\left[\gamma^\mu\gamma^\nu\right]^\rho_{34}
\left[\gamma_\mu\right]^\sigma_{12} \left[\gamma_\nu\right]^\tau_{56}$
and terms that are free from double contractions such as $\gamma^\mu
\gamma^\nu \otimes \gamma_\mu \otimes \gamma_\nu$.

{\it Step 1:} If $n_i>1$ for $i=1$ or 2, then we write $A_i=\tilde A_i
\ps_a\ps_b$ and, using \refeq{eq:chisholm1}, we shift $\ps_a\ps_b$ to
the chain that contains $A_0\gamma^\mu\gamma^\nu$.  Then we perform
the simplifications (i)--(iii) in four dimensions.  This step is
iterated until $n_1,n_2\le 1$.

{\it Step 2:} If $n_1,n_2\le 1$ and $n_0>0$, then we can write
$A_0=\tilde A_0 \ps_a$ with $a\in \{k,l,m,n\}$, since $\ps_{i,j}$ are
eliminated by means of (i)--(iii), also in four dimensions.  In this
case, using \refeq{eq:chisholm1}, we shift $\ps_a$ and one of the
matrices $\gamma^\mu,\gamma^\nu$ from the $A_0$-chain to that
$A_{i}$-chain where we can eliminate $\ps_a$ by means of the Dirac
equation and other simplifications (i)--(iii). Then we restart with
step 1.

This procedure recursively reduces the number of $\ps$-terms
$n_0+n_1+n_2$ until $n_1,n_2\le 1$ and $n_0=0$, which automatically
implies $n_0=n_1=n_2=0$ since only one of the three Dirac chains (the
massive one) can contain an even number of Dirac matrices.
Thus the only $ \gamma^\mu \gamma^\nu \otimes \gamma_\mu \otimes
\gamma_\nu $ structure that survives is
$\left[\gamma^\mu\gamma^\nu\right]^\rho_{34}
\left[\gamma_\mu\right]^\sigma_{12} \left[\gamma_\nu\right]^\tau_{56}
$.
\item[] {\it Example 2:} \\
We consider Dirac structures of the type
$\left[\gamma^\mu\right]^\rho_{12}
\left[\gamma_\mu\dsl{p}_k\right]^\sigma_{34}
\left[\dsl{p}_l\right]^\tau_{56}$.
Using the relations
\beqar
\left[\gamma^\mu\right]^\pm_{12} \left[\dsl{p}_1\gamma_\mu \right]^\pm_{34}
&=&
-\frac{1}{m_3} \left[\gamma^\mu\right]^\pm_{12} 
\left[\dsl{p}_3\dsl{p}_1\gamma_\mu \right]^\pm_{34}
= -\frac{1}{m_3} \left[\dsl{p}_1\dsl{p}_3\gamma^\mu\right]^\pm_{12}
\left[\gamma_\mu\right]^\pm_{34} = 0,
\nn\\
\left[\gamma^\mu\right]^\pm_{12} \left[\gamma_\mu\dsl{p}_2\right]^\mp_{34}
&=&
+\frac{1}{m_4} \left[\gamma^\mu\right]^\pm_{12} 
\left[\gamma_\mu\dsl{p}_2\dsl{p}_4\right]^\pm_{34}
= +\frac{1}{m_4} \left[\gamma^\mu\dsl{p}_4\dsl{p}_2\right]^\pm_{12} 
\left[\gamma_\mu\right]^\pm_{34} = 0,
\eeqar
which follow from \refeq{eq:chisholm1}, and using momentum
conservation, we can achieve that the index $k$ takes only the values
$k=5,6$ for each chirality configuration $(\rho\sigma\tau)$.
Eliminating one $p_l$ via momentum conservation, the index $l$ can
take three values, leading to six different index pairs $(kl)$ per
chirality configuration.  Two out of the six possibilities can be
easily eliminated by relations like \refeq{eq:chisholm3}:
\beqar\label{exampletwo}
\left[\gamma_\mu \dsl{p}_5\right]^\pm_{34} \left[\dsl{p}_4\right]^\pm_{56} 
&=&
(p_4 p_5) \left[\gamma_\mu\gamma_\nu\right]^\pm_{34} 
\left[\gamma^\nu\right]^\pm_{56} 
-\frac{m_4}{2} \left[\gamma_\mu\gamma_\nu\dsl{p}_5\right]^\mp_{34} 
\left[\gamma^\nu\right]^\pm_{56},
\nn\\
\left[\gamma_\mu \dsl{p}_6\right]^\pm_{34} \left[\dsl{p}_4\right]^\mp_{56} 
&=&
(p_4 p_6) \left[\gamma_\mu\gamma_\nu\right]^\pm_{34} 
\left[\gamma^\nu\right]^\mp_{56} 
-\frac{m_4}{2} \left[\gamma_\mu\gamma_\nu\dsl{p}_6\right]^\mp_{34} 
\left[\gamma^\nu\right]^\mp_{56},
\nn\\
\left[\gamma_\mu \dsl{p}_5\right]^\pm_{34} \left[\dsl{p}_3\right]^\pm_{56} 
&=&
-(p_3 p_5) \left[\gamma_\nu\gamma_\mu\right]^\pm_{34} 
\left[\gamma^\nu\right]^\pm_{56} 
-\frac{m_3}{2} \left[\dsl{p}_5\gamma_\nu\gamma_\mu\right]^\pm_{34} 
\left[\gamma^\nu\right]^\pm_{56}
+2p_{5,\mu} \left[ 1\right]^\pm_{34}
\left[\dsl{p}_3\right]^\pm_{56},
\nn\\
\left[\gamma_\mu \dsl{p}_6\right]^\pm_{34} \left[\dsl{p}_3\right]^\mp_{56} 
&=&
-(p_3 p_6) \left[\gamma_\nu\gamma_\mu\right]^\pm_{34} 
\left[\gamma^\nu\right]^\mp_{56} 
-\frac{m_3}{2} \left[\dsl{p}_6\gamma_\nu\gamma_\mu\right]^\pm_{34} 
\left[\gamma^\nu\right]^\mp_{56}
+2p_{6,\mu} \left[ 1\right]^\pm_{34}
\left[\dsl{p}_3\right]^\mp_{56},
\nn\\
\eeqar
where the use of \refeq{eq:chisholm3} in the last two equations
required an anticommutation of \mbox{$\gamma_\mu \, \dsl{p}_{5,6}$}
leading to additional contributions, and the terms proportional to
$m_{3,4}$ on the right-hand side of \refeq{exampletwo} can be further
reduced as in Example 1.  Another $(kl)$ combination can be eliminated
by using identities like \refeq{eq:chisholm3} for the chains
$[\dots]_{12}$ and $[\dots]_{56}$ after shifting $\dsl{p}_k$ to
$[\dots]_{12}$. In order to achieve this, one has to apply the Dirac
equation for the massive fermion inversely in the first step:
\beqar
\left[\gamma_\mu\right]^\pm_{12}
\left[\gamma^\mu \dsl{p}_6\right]^\pm_{34} 
\left[\dsl{p}_1\right]^\pm_{56} 
&=&
\frac{1}{m_4}
\left[\gamma_\mu\right]^\pm_{12}
\left[\gamma^\mu \dsl{p}_6\dsl{p}_4\right]^\mp_{34} 
\left[\dsl{p}_1\right]^\pm_{56} 
= \frac{1}{m_4}
\left[\dsl{p}_6\dsl{p}_4\gamma_\mu\right]^\pm_{12}
\left[\gamma^\mu\right]^\mp_{34}
\left[\dsl{p}_1\right]^\pm_{56}
\nn\\
&=& 
\frac{(p_1 p_6)}{m_4}
\left[\gamma_\nu\dsl{p}_4\gamma_\mu\right]^\pm_{12}
\left[\gamma^\mu\right]^\mp_{34}
\left[\gamma^\nu\right]^\pm_{56}
= 
\frac{(p_1 p_6)}{m_4}
\left[\gamma_\mu\right]^\pm_{12}
\left[\gamma^\mu\gamma_\nu\dsl{p}_4\right]^\mp_{34}
\left[\gamma^\nu\right]^\pm_{56}
\nn\\
&=&
(p_1 p_6) \left[\gamma_\mu\right]^\pm_{12}
\left[\gamma^\mu\gamma_\nu\right]^\pm_{34}
\left[\gamma^\nu\right]^\pm_{56},
\nn\\
\left[\gamma_\mu\right]^\pm_{12}
\left[\gamma^\mu \dsl{p}_5\right]^\pm_{34} 
\left[\dsl{p}_1\right]^\mp_{56} 
&=&
\dots = 
(p_1 p_5) \left[\gamma_\mu\right]^\pm_{12}
\left[\gamma^\mu\gamma_\nu\right]^\pm_{34}
\left[\gamma^\nu\right]^\mp_{56},
\nn\\
\left[\gamma_\mu\right]^\pm_{12}
\left[\gamma^\mu \dsl{p}_6\right]^\mp_{34} 
\left[\dsl{p}_2\right]^\mp_{56} 
&=&
\dots = 
(p_2 p_6) \left[\gamma_\mu\right]^\pm_{12}
\left[\gamma^\mu\gamma_\nu\right]^\mp_{34}
\left[\gamma^\nu\right]^\mp_{56},
\nn\\
\left[\gamma_\mu\right]^\pm_{12}
\left[\gamma^\mu \dsl{p}_5\right]^\mp_{34} 
\left[\dsl{p}_2\right]^\pm_{56} 
&=&
\dots = 
(p_2 p_5) \left[\gamma_\mu\right]^\pm_{12}
\left[\gamma^\mu\gamma_\nu\right]^\mp_{34}
\left[\gamma^\nu\right]^\pm_{56}.
\eeqar
One additional relation per chirality configuration results upon
exploiting $0=
[\gamma^\mu(\dsl{p}_1+\dots+\dsl{p}_6)\gamma^\nu]^\rho_{12}
[\gamma_\mu\dsl{p}_k]^\sigma_{34} [\gamma_\nu]^\tau_{56}$ similar to
Step~5 in Section~3.3 of \citere{Denner:2005fg}, however, this
procedure is quite tedious.
\end{itemize}
The complete reduction algorithm consists of several procedures of
this type, each consisting of combinations of the identities
\refeq{eq:chisholm1}--\refeq{eq:chisholm3} and the operations
(i)--(iii).

In the case of massless 6-fermion processes \cite{Denner:2005fg}, all
Dirac structures were reduced to 10 types of SMEs of the form
\beqar\label{masslessSMEs}
\left[\gamma^\mu\right]^\rho_{ij}
\left[\gamma^\mu\right]^\sigma_{kl}
\left[\ps_a\right]^\tau_{mn}
,\qquad
\left[\ps_a\right]^\rho_{ij}
\left[\ps_b\right]^\sigma_{kl}
\left[\ps_c\right]^\tau_{mn}.
\eeqar
Counting the different chiralities $\rho,\sigma,\tau=\pm$, which yield
8 or less combinations per type of SME depending on the type, the
total number of independent ``massless'' SMEs was 80.  In addition to
these SMEs, the $q\bar q \to \Pt\bar\Pt\Pb\bar\Pb$ reduction yields 15
types of SMEs of the form
\beqar\label{massiveSMEsa}
\left[\gamma^\mu\right]^\rho_{ij}
\left[\gamma_\mu\right]^\sigma_{kl}
\left[1\right]^\tau_{34}
,\quad
\left[\gamma^\mu\right]^\rho_{ij}
\left[\gamma^\nu\right]^\sigma_{kl}
\left[\gamma_\mu\gamma_\nu\right]^\tau_{34}
,\quad
\left[\ps_a\right]^\rho_{ij}
\left[\gamma^\mu\right]^\sigma_{kl}
\left[\gamma_\mu\ps_b\right]^\tau_{34}
,\quad
\left[\ps_a\right]^\rho_{ij}
\left[\ps_b\right]^\sigma_{kl}
\left[1\right]^\tau_{34}
,
\hspace{2em}
\eeqar
where the chain $[\dots]_{34}$, \ie the top-quark chain, involves an
even number (0 or 2) of Dirac matrices.  In the two independent
reduction algorithms that we have implemented the total number of SMEs
for $q\bar q \to \Pt\bar\Pt\Pb\bar\Pb$, counting all types
\refeq{masslessSMEs}--\refeq{massiveSMEsa} and chiralities
$\rho,\sigma,\tau$, is 148 and 156.  As it is obvious, the presence of
the top mass increases the number of the SMEs by roughly a factor 2.
Moreover, also the complexity of the form factors associated with each
SME grows considerably with respect to the case where all fermions are
massless.

\section{Benchmark numbers for the virtual corrections}
\label{app:benchmark}

In order to facilitate a comparison to our calculation, in this
appendix we provide explicit numbers on the 
squared LO amplitude and the corresponding virtual correction for a
single non-exceptional phase-space point. The set of momenta for
the partonic reaction $q\bar q \to \Pt\bar\Pt\Pb\bar\Pb$ is chosen as
\beqar
p_q^\mu &=& \scriptstyle (500,0,0,500),
\nn\\
p_{\bar q}^\mu &=& \scriptstyle (500,0,0,-500),
\nn\\
p_{\Pt}^\mu &=& \scriptstyle
(327.5045589027869,
 107.1276753641986,
-107.9290580423663,
-233.1168284428635),
\nn\\
p_{\bar\Pt}^\mu &=& \scriptstyle
(276.6425142763093,
-107.4949148022111,
 153.8289259355409,
-107.3397668261919),
\nn\\
p_{\Pb}^\mu &=& \scriptstyle
(233.9459027189062,
  82.55875671042013,
 -77.70592645955253,
 204.6375480757531),
\nn\\
p_{\bar\Pb}^\mu &=& \scriptstyle
(161.9070241019976,
 -82.19151727240762,
  31.80605856637796,
 135.8190471933023),
\label{eq:PSpoint}
\eeqar
with the components given in GeV and $\Mt=172.6\GeV$.  We give numbers
on the spin- and colour-averaged squared LO amplitude $|\M^{\LO}|^2$
and on the sum of the relative virtual NLO correction $\de_{\virt}$
and the contribution $\de_{\rI}$ of the $I$ operator of the dipole
subtraction function as defined in \citere{Catani:2002hc}.  In more
detail, we split the relative correction into a contribution
originating from closed fermion loops, $\de_{\ferm}$ (comprising
contributions from the gluon self-energy, the triple-gluon vertex
correction, and the renormalization constant of the strong coupling),
and the remaining loop corrections, called $\de_{\bos}$, and
$\de_{\rI}$.  Note that for $q\bar q \to \Pt\bar\Pt\Pb\bar\Pb$ the
fermionic part is IR finite, while $\de_{\bos}$ is IR divergent.
Adding $\de_{\rI}$ to $\de_{\bos}$ or
$\de_{\virt}=\de_{\ferm}+\de_{\bos}$, all IR divergences cancel, and
the sum is independent of the IR regularization scheme.  The values of
the strong coupling constant at $\mu_{\mathrm{R}}=\Mt$ in the setup
described in \refse{se:numres} are
\beq
\alpha_{\mathrm{s}}(\Mt)|_{\mathrm{LO}} =0.1178730139006150, \qquad
\alpha_{\mathrm{s}}(\Mt)|_{\mathrm{NLO}}=0.1076396017050965.
\eeq
At the phase-space point \refeq{eq:PSpoint} we find
\beqar
|\M^{\LO}|^2/g_{\mathrm{s}}^8 &=&  0.4487410759198035\cdot10^{-8}\GeV^{-4},
\nn\\
|\M^{\LO}|^2/g_{\mathrm{s}}^8\Big|_{\mathrm{Madgraph}} 
&=&  0.4487410759198011\cdot10^{-8}\GeV^{-4},
\nn\\
\de_{\virt+\rI}\Big|_{\mathrm{version 1}} &=& -0.1290522911043483,
\nn\\
\de_{\virt+\rI}\Big|_{\mathrm{version 2}} &=& -0.1290522911137204,
\nn\\
\de_{\ferm} \Big|_{\mathrm{version 1}} &=&   -0.06326213639716407,
\nn\\
\de_{\ferm} \Big|_{\mathrm{version 2}} &=&   -0.06326213639715421,
\nn\\
\de_{\bos+\rI}\Big|_{\mathrm{version 1}} &=& -0.06579015470718421,
\nn\\
\de_{\bos+\rI}\Big|_{\mathrm{version 2}} &=& -0.06579015471656619,
\eeqar
where we divided out the strong coupling constant $g_{\mathrm{s}}$
from $|\M^{\LO}|^2$.  The agreement between our two independent
versions of the virtual corrections is typically about 10~digits at
regular phase-space points.

\end{document}